\documentclass[a4paper,12pt]{article}
\makeatletter

\usepackage[english]{babel}
\usepackage{lmodern}
\usepackage{microtype}

\usepackage{mathtools}
\usepackage{fullpage, euscript, amsfonts, amssymb, amsmath}
\usepackage{diffcoeff}
\usepackage[cal=boondox]{mathalfa}
\usepackage{lmodern,textcomp}
\usepackage{pifont}
\usepackage{accents}
\newcommand{\ubar}[1]{\underaccent{\bar}{#1}}

\usepackage{setspace}
\usepackage[paper=a4paper, 
            inner=22mm, 
            outer=22mm, 
            top=22mm, 
            bottom=30mm, 
            headheight=6mm,
            includehead,
            includefoot, 
            heightrounded, 
            twoside]{geometry}
            
\usepackage{lscape}
\usepackage{placeins}
\usepackage{ragged2e}

\usepackage{graphicx}
\usepackage{float}
\usepackage{subcaption}
\usepackage{import}

\usepackage{tabularx}
\usepackage{booktabs}
\usepackage{threeparttable}
\usepackage{multirow}
\usepackage{hhline}
\usepackage{adjustbox}
\usepackage{siunitx}

\usepackage[
    backend=biber,
    style=authoryear,
    sorting=nyt,
    url=false,
    dashed=false
]{biblatex}
\usepackage{csquotes}
\DeclareFieldFormat[techreport, report, misc]{title}{\mkbibquote{#1}}
\DeclareFieldFormat[techreport, report]{type}{\mkbibemph{#1}}
\DeclareFieldFormat{eprint}{\texttt{#1}}

\DeclareNameAlias{sortname}{family-given}

\AtBeginBibliography{\defcounter{maxnames}{99}} 

\DeclareBibliographyDriver{*}{%
  \usebibmacro{bibindex}%
  \usebibmacro{begentry}%
  \printnames{author}%
  \setunit{\addcomma\space}%
  \printfield{year}%
  \setunit{\addcomma\space}%
  \printfield{title}%
  \newunit\newblock
  \printfield{journaltitle}%
  \setunit{\addcomma\space}%
  \printfield{volume}%
  \setunit{\addcomma\space}%
  \printfield{number}%
  \setunit{\addcomma\space}%
  \printfield{pages}%
  \newunit\newblock
  \printfield{institution}%
  \setunit{\addcomma\space}%
  \printfield{type}%
  \setunit{\addcomma\space}%
  \printfield{number}%
  \usebibmacro{finentry}%
}

\DeclareBibliographyDriver{article}{%
  \usebibmacro{bibindex}%
  \usebibmacro{begentry}%
  \printnames{author}%
  \setunit{\addspace}%
  \printtext[parens]{\printfield{year}}%
  \setunit{\addperiod\space}%
  \printfield{title}%
  \newunit\newblock
  \printfield{journaltitle}%
  \setunit{\addcomma\space}%
  \printfield{volume}%
  \setunit{\addcomma\space}%
  \printfield{number}%
  \setunit{\addcomma\space}%
  \printfield{pages}%
  \usebibmacro{finentry}%
}

\usepackage{multirow}
\usepackage{xcolor}
\usepackage{xurl}
\definecolor{my_blue}{HTML}{1f77b4}

\usepackage{appendix}

\usepackage{bigfoot}
\usepackage{caption}

\usepackage[
    pdftitle={Households Financial Inclusion},
    pdfauthor={Maria Elena Filippin},
    hidelinks,
    hyperfootnotes=false
]{hyperref}
\hypersetup{
  colorlinks,
  citecolor=black,
  linkcolor=black}

\newcommand\blfootnote[1]{%
  \begingroup
  \renewcommand\thefootnote{}\footnote{#1}%
  \addtocounter{footnote}{-1}%
  \endgroup
}

\makeatother
\title{Who's in? Household-targeted Government Policies and the Role of Financial Literacy in Market Participation}

\DeclareNewFootnote{AAffil}[arabic]
\DeclareNewFootnote{ANote}[fnsymbol]

\usepackage{etoolbox}
\makeatletter
\patchcmd\maketitle{\def\@makefnmark{\rlap{\@textsuperscript{\normalfont\@thefnmark}}}}{}{}{}

\makeatletter
\def\thanksAAffil#1{
  \footnotemarkAAffil\protected@xdef\@thanks{\@thanks%
        \protect\footnotetextAAffil[\the \c@footnoteAAffil]{#1}}%
}

\def\thanksANote#1{%
  \footnotemarkANote%
  \protected@xdef\@thanks{\@thanks%
        \protect\footnotetextANote[\the \c@footnoteANote]{#1}}%
}

\makeatother

\author{%
  Maria Elena Filippin%
  \thanksANote{Central Bank of Ireland and Uppsala University. \textit{E-mail}: \texttt{mariaelena.filippin@centralbank.ie}}%
}

\date{February 19, 2026}

\begin{document}

\newgeometry{top=0.95in, bottom=0.95in, left=22mm, right=22mm}
\maketitle
\thispagestyle{empty}
\maketitle

\begin{abstract}
This paper examines how household-targeted government policies influence financial market participation conditional on financial literacy, focusing on potential Central Bank Digital Currency (CBDC) adoption. Due to the lack of empirical CBDC data, I use the 2012 introduction of retail Treasury bonds in Italy as a proxy to study how financial literacy affects households' likelihood to engage with a new government-backed retail instrument. Using the Bank of Italy's Survey on Household Income and Wealth, I show that households with some but low financial literacy are more likely to participate in the Treasury bond market than other groups following the introduction of the new instrument. Based on these findings, I develop a theoretical model to study how financial literacy affects CBDC demand through portfolio choice: low-literate households with limited access to risky assets allocate more wealth to CBDC, while high-literate households use risky assets to safeguard against income risk. These results highlight the role of financial literacy in shaping portfolio choices and CBDC adoption.

\vspace{0.5cm}
\textbf{Keywords}: Central Bank Digital Currency, Financial literacy, Government policies, Market participation, Treasury bonds\blfootnote{I am grateful to Ulf Söderström and Karl Walentin for their invaluable guidance and continuous support. For their useful feedback and comments, I also thank Nicola Branzoli, Peter Fredriksson, Georg Graetz, David Loschiavo, Dirk Niepelt (discussant), Roine Vestman (discussant), as well as conference and seminar participants at Uppsala University, Sveriges Riksbank, Banca d'Italia, Central Bank of Ireland, and the WIRE First Annual Workshop. I am also grateful to the anonymous referee for their constructive comments, which helped improve the paper. A large part of this work was conducted during my PhD at Uppsala University. The views expressed here are solely my own and do not necessarily reflect the views of the Central Bank of Ireland.}
\end{abstract}
\clearpage
\setcounter{page}{1}
\restoregeometry

\section*{Non-Technical Summary}
This paper investigates how household-targeted government policies affect market participation conditional on financial literacy and provides intuition on how demand for a Central Bank Digital Currency (CBDC) may vary across households. Central banks worldwide are still assessing whether and how to issue a CBDC --- a digital form of central bank money. Given the lack of empirical CBDC data, I use a proxy policy to study how financial literacy affects households' likelihood of engaging with the new financial instrument.

The proxy is the introduction of retail Treasury bonds in Italy. While not perfectly equivalent, both instruments are government-backed and household-oriented. Retail Treasury bonds --— a subcategory of Treasury bonds exclusively issued to households --- are investment instruments, whereas many central banks envisage CBDC as a means of payment. Here, CBDC is considered useful for transactional and store-of-value purposes, following \textcite{Li2023}.

I use data from the Bank of Italy’s Survey on Household Income and Wealth to empirically assess the impact of the first retail Treasury bond \textit{Italia} on participation in the broader Treasury bond market. Introduced in March 2012 with yields indexed to domestic inflation, the Treasury bond \textit{Italia} offered an ex-ante attractive investment compared to a comparable standard Treasury Bond, given positive inflation expectations. Results show that financial literacy increases the propensity to participate after introducing the new instrument, but the effect varies across literacy levels: households with some but low literacy are more likely to participate than financially illiterate and highly literate households. The findings also suggest a portfolio reallocation from other government securities and stocks toward Treasury bonds.

Building on these insights, I develop a simple two-period model with two types of agents differentiated by financial literacy, which determines access to risky investments. The high-literacy agent can invest in risky assets and risk-free deposits, while the low-literacy agent is limited to deposits. Agents live for two periods, receive an endowment in the first period, and an income in the second. They derive utility from consumption and value liquidity services from holding financial assets. In the economy with a retail CBDC, the low-literacy agent allocates more wealth to CBDC, irrespective of second-period income values. This reflects limited access to risky assets, while the high-literacy agent adjusts the portfolio by balancing risk and liquidity needs. When the second-period income is uncertain, differences between agents reduce: as uncertainty increases, the high-literacy agent shifts from risky assets toward safer options like CBDC.

Overall, the findings show that households with different levels of financial literacy respond differently to new financial instruments. For CBDC, clear communication of its benefits and functionality is essential to foster broad and inclusive adoption.

\section{Introduction}

It has been well-documented that household participation in financial markets is limited, resulting in potential welfare costs due to non-participation [\textcite{Guiso2013}, \textcite{Haliassos2021}]. Government policies targeted at households have the potential to increase market participation. A particularly relevant policy instrument that central banks worldwide are considering is the issuance of public money in digital form, known as Central Bank Digital Currency (CBDC). A retail CBDC would be accessible to the general public, allowing everyone to have a digital wallet at a financial institution. Central banks discuss CBDCs as a way to preserve access to public money as cash use declines and to maintain monetary policy effectiveness in an increasingly digital economy, with some designs also aiming to support financial inclusion and improve cross-border payments. Importantly, CBDC proposals are intended to coexist with cash, as a digital complement to it.

While most research on CBDCs focuses on implications for the banking system and financial stability, there is little evidence of the effects of introducing CBDC on market participation. This lack arises because central banks are still assessing whether and how to issue a CBDC, making it difficult to empirically investigate adoption patterns.\footnote{
A handful of central banks have already issued retail CBDCs; see \textcite{Branch2025} for insights from the practical experiences with the Sanddollar, E-CNY and JAM-DEX®.
}
Although everyone could access CBDC, the extent to which households will use it might differ based on personal characteristics and preferences. While CBDC would be relatively easy to use in principle, adoption may depend on individuals' understanding of its benefits and functionality, which likely depends on their level of financial literacy. CBDC use would be voluntary, but very low adoption would limit CBDC's ability to deliver its policy objectives.

This paper provides intuition on how CBDC demand might vary based on household characteristics. Specifically, it evaluates the effects of financial literacy on CBDC adoption. Given the lack of available empirical data on CBDC, I use a proxy to explore how financial literacy affects households' likelihood of engaging with the new instrument.

The proxy is the introduction of retail Treasury bonds in Italy. Both instruments are claims on the government and are household-oriented. Retail Treasury bonds are investment instruments, whereas many central banks envision CBDC primarily as a means of payment. I use the Treasury bond episode for a narrower purpose: to study the household adoption problem that policymakers face when introducing a new government-issued retail instrument. In the retail Treasury bond episode, households access the new instrument through an institutional channel and evaluate it relative to alternative assets. Similarly, a retail CBDC would be accessed through a digital wallet, and adoption would likely depend on households’ assessment of how CBDC compares to existing liquid assets. In both cases, assessment of the instruments’ functionality may vary across levels of financial literacy. The proxy, therefore, speaks to heterogeneity in adoption when a new public retail instrument is introduced, rather than to payment-specific forces such as merchant acceptance or network effects.

I consider a CBDC useful for both transactional and store-of-value purposes, similar to \textcite{Li2023}. In the theoretical framework, I model CBDC as interest-bearing but return-dominated by deposits. Although central banks are currently considering non-interest-bearing CBDC designs, I later show that the results are robust even in this case. All in all, while not a perfect substitute, retail Treasury bonds offer a suitable comparative proxy for studying CBDC adoption. Based on the empirical findings, I then develop a simple illustrative model to explore how CBDC demand differs by household financial literacy.

The first part of the paper uses data from the Bank of Italy's Survey on Household Income and Wealth (SHIW) to assess the impact of introducing Italian government bonds tailored to household investors on financial market participation. Italy presents an interesting case, as it has one of the lowest levels of financial literacy among OECD countries [\textcite{OECD2020}], while also showing one of the highest amounts of households' direct holdings of government debt among euro area countries [\textcite{Pavot2021}]. These facts make Italy a particularly relevant setting for studying Treasury bond market participation when considering financial literacy.

In 2012, the Italian government introduced retail Treasury bonds, a subcategory of Treasury bonds exclusively issued to retail investors (i.e., households) through a dedicated section of the Milan Stock Exchange. These bonds can be purchased through a home banking system with an online trading feature, allowing households to buy bonds without visiting a bank.\footnote{
Additionally, buyers during the placement period are guaranteed to receive the quantity they request.}
This direct purchase mechanism likely requires some degree of financial literacy, which is accounted for in the analysis.

I empirically evaluate the effects of the introduction of retail Treasury bonds on household participation, conditional on financial literacy. Specifically, the empirical exercise addresses several key questions: (i) did financial literacy influence households' participation in the Treasury bond market following the introduction of retail Treasury bonds?, (ii) how did the effect vary across financial literacy levels?, and (iii) did households reallocate their portfolios after the introduction of retail Treasury bonds? Importantly, the focus is on the extensive margin of participation (i.e., who started to hold Treasury bonds) rather than the intensive one (i.e., how much Treasury bonds are held).

The empirical results reveal that financial literacy positively influenced participation in the Treasury bond market after the introduction of retail Treasury bonds. Furthermore, the effect is non-monotone across literacy levels, with households with some but low financial literacy being more likely to participate following the introduction of the new instrument compared to the financially illiterate and highly literate household groups. The findings also suggest a reallocation of portfolios, with households shifting investments from other government securities and stocks toward Treasury bonds.

The second part of the paper explores the potential implications of financial literacy for CBDC adoption. I use the empirical insights to inform the theoretical framework, which considers a simple two-period endowment economy with two types of agents, differentiated by their financial literacy, which determines access to risky investments. The high-financially literate agent can invest in risky assets and risk-free deposits, while the low-financially literate agent is limited to deposits. Each agent lives for two periods, receives an initial endowment in the first period, and an income in the second. Agents derive utility from consumption and value liquidity services from holding financial assets. The government introduces a CBDC accessible to both types of agents, which is potentially interest-bearing and offers benefits relative to deposits.

The model examines how high- and low-financially literate agents allocate their portfolios in different economic environments. If second-period income is deterministic, the results reveal that low-financially literate households hold more CBDC than high-financially literate households, who rely on risky assets to safeguard against income risk, thereby reducing their need for CBDC. Introducing stochastic second-period income reduces these differences: as uncertainty rises, high-financially literate households reallocate away from risky assets toward safer options like CBDC. This analysis offers insights relevant to policymakers for designing an effective CBDC that can achieve broad adoption across households with different financial literacy levels.\\

\noindent 
\textbf{Related literature}.  \hspace{0.5mm} This paper contributes to the diverse literature on financial literacy and market participation, encompassing studies from various countries. \textcite{Arrondel2016} investigate the relationship between financial literacy and financial behavior in the French population using the 2011 PATER household survey and find that financial literacy significantly impacts the probability of holding stocks. Similarly, using the 2005 De Nederlandsche Bank's Household Survey, \textcite{Vanrooij2011} highlight financial literacy as a key determinant of stock market participation. In recent work, \textcite{ChenDai2023} use data from the China Household Finance Survey to demonstrate that higher levels of advanced financial literacy substantially increase the propensity of stock market participation, with an additional point in the advanced financial literacy score raising the probability by approximately 12 percentage points. \textcite{Hsiao2018} utilize data from the 2011 National Financial Literacy Surveys in Taiwan and find that individuals with higher financial literacy levels are more likely to purchase derivative products.

In Italy, studies have revealed low levels of financial literacy. Using the Italian Literacy and Financial Competence Survey (IACOFI), conducted by the Bank of Italy in early 2017, \textcite{Disalvatore2018} highlight a significant gap in financial literacy between Italy and other G20 countries, particularly among less educated individuals, the elderly, and women. Similarly, using the second IACOFI in 2020, \textcite{Dalessio2020} document that the financial literacy of Italians lags behind by international standards and varies across different population groups. They confirm their results using some waves of the SHIW data. Their findings are consistent with earlier evidence from \textcite{Guiso2005} who, using the 1995 and 1998 SHIW, show that non-awareness of financial products helps explain the stock-holding puzzle. They observe limited direct stock holding among Italian households during that period, with market participation increasing over time and being correlated with household resources, education levels, and geographical characteristics. \textcite{Gallo2023} use the 2016 SHIW and show that financial literacy influences the values and inequality levels of Italian household income and wealth.

This paper also contributes to the fast-developing literature on CBDC. Most studies focus on the implications of CBDC for the banking systems and financial stability [see, e.g., \textcite{Assenmacher2021}, \textcite{Burlon2022}, \textcite{Chen2024}, \textcite{Chiu2023}, \textcite{Whited2023}, and \textcite{Williamson2022}]. A small strand studies potential CBDC demand. An exception is \textcite{Li2023} who investigates household demand for CBDC in Canada. Using a structural demand model applied to survey data, she estimates that CBDC could account for $4\%$ to $52\%$ of household liquid assets, depending on how it compares to cash and deposits in terms of features like anonymity, usefulness for budgeting, and bundling with banking services. Moreover, her study highlights that allowing banks to respond to CBDC could significantly constrain its adoption.

Using survey data from the Netherlands, \textcite{Bijlsma2021} analyze public adoption intentions for CBDC. Their findings suggest that consumers perceive CBDC as distinct from traditional bank accounts, with privacy, security, and trust in the central bank playing crucial roles in driving adoption. The study also suggests that interest rates and the design of CBDC could influence public adoption and usage. In another work, \textcite{huynh2020} develop a structural model of demand for payment instruments and explore how consumers might use CBDC for point-of-sale (POS) payments, examining how payment method attributes influence consumer choices. Similarly, \textcite{Nocciola2024} assess transactional demand for CBDC at the POS, using a structural model of payment adoption and usage. They focus on the frictions associated with CBDC adoption, such as information barriers and the gradual diffusion of digital payment methods, and emphasize the importance of optimal CBDC design, information campaigns, and network effects in boosting CBDC demand. 

This paper contributes to both literatures. First, it provides evidence on the effect of a government's household-oriented bond policy on Treasury bond market participation and portfolio allocation, and on how the effect varies across levels of financial literacy. Second, it complements the CBDC demand literature by using the empirical insights to discipline an illustrative model of CBDC demand and portfolio choice under heterogeneity in financial literacy. While \textcite{huynh2020} and \textcite{Nocciola2024} focus on CBDC as a means of payment, this study considers a CBDC useful for both transactional and store-of-value purposes, similar to \textcite{Li2023}. The model also examines how agents with different levels of financial literacy adjust their portfolio choices, including CBDC demand, when facing uncertainty.\\

\noindent
The rest of the paper is organized as follows. Section \ref{itbonds} provides an overview of the Italian government bond market. Section \ref{data} describes the data used in the empirical exercise. Section \ref{empexercise} presents the empirical analysis and its results. Section \ref{theory} describes the theoretical framework and portfolio choices following the CBDC introduction. Section \ref{conclusion} concludes.

\section{Italian government bond market} \label{itbonds}
This section outlines the Italian government bond market. The landscape of Italian government securities is diverse, with four broad categories available to private and institutional investors: Treasury bills, Zero-coupon bonds, Treasury certificates, and Treasury bonds.\footnote{
Table \ref{bondcat} in Appendix \ref{sec:app:additional} summarizes the differences among government securities in terms of maturity, remuneration, and auction type.}
This work focuses on Treasury bonds, i.e., in Italian, \textit{Buoni del Tesoro Poliennali} (\textit{BTPs}), which translates to Treasury bonds with long-term maturity. Treasury bonds are a core part of Italy's public debt structure. As of the end of 2011, they represented $66.47\%$ of the total Italian government securities market [\textcite{mef2011}].

To boost domestic holding of its debt, in March 2012 the Italian government introduced retail Treasury bonds which are available to retail investors only. The first retail Treasury bonds were named \textit{Italia}, followed by \textit{Futura} in 2020 and \textit{Valore} in 2023. Besides being available exclusively to retail investors, retail Treasury bonds differ from standard ones in that they offer a final bonus to whoever purchases them during the placement period and holds them until maturity. Treasury bonds \textit{Italia} have yields linked to the Italian Consumer Price Index, while Treasury bonds \textit{Futura} and \textit{Valore} provide nominal coupons increasing over time, calculated based on a preset path of increasing rates over time, i.e., a ``step-up'' mechanism. Table \ref{btps} reports the characteristics of both standard and retail Treasury bonds.

\begin{table}[!htbp]
\centering
\caption{Treasury bonds (\textit{BTPs}) in Italy}
\begin{threeparttable}
\begin{tabular}{l p{2.33cm} p{3.33cm} p{3.33cm} p{3.33cm}}
\toprule\toprule
& Standard & \multicolumn{3}{c}{Retail} \\
\cmidrule(lr){3-5}
& & \textit{Italia} & \textit{Futura} & \textit{Valore}\\
\midrule
Maturity & 3-50 years & 4-8 years & 8-16 years & 4-6 years \\
Coupon & Floating semi-annual & Semi-annual & Fixed semi-annual & Fixed \\
Yield & Avg. $3.86\%$ &  Min. $2\%$ indexed to inflation & ``Step-up" mechanism: $1.15\%$ ($1.3\%$) [$1.45\%$] for the first 4y (middle 3y) [last 2y] & ``Step-up" mechanism: $3.25\%$ ($4\%$) for the first 3y (last 3y) \\
Final bonus & No & Yes & Yes* & Yes \\
\bottomrule
\end{tabular}
\end{threeparttable}
\vskip 0.5em
\begin{minipage}{\textwidth}
    \footnotesize
    \textit{Note}: This table compares the types of retail Treasury bonds among themselves and with the standard Treasury bonds. The standard Treasury bond yield refers to the year 2023 and represents the weighted average interest rate. All Treasury bonds have a minimum denomination of €$1,000$ and are subject to a $12.5\%$ tax rate. * The \textit{Futura} final bonus is linked to GDP. \\
    \textit{Source}: Italian Ministry of Economy and Finance.
\end{minipage}
\label{btps}
\end{table}

The issuance of retail Treasury bonds is primarily driven by the government's financing needs and strategic debt management plans, rather than direct demand from retail investors. The government schedules retail Treasury bond offerings to align with its budgetary requirements and broader economic objectives. For example, the introduction of Treasury bonds \textit{Futura} and \textit{Valore} was part of efforts to support economic recovery and public debt management following the 2019 COVID crisis.

Retail Treasury bonds are allocated through a non-competitive subscription process during predefined offering periods, ensuring all valid orders are fully satisfied. The Italian Ministry of Economy and Finance (MEF) announces these offerings in advance and provides detailed terms and conditions to inform potential investors, while the bonds are issued via the MOT platform of the Milan Stock Exchange to guarantee broad accessibility for individual investors. Since their introduction, retail Treasury bonds have received a positive response from retail investors. For instance, the most recent ones, \textit{Valore}, saw a constant positive take-up since their first issuance. See Appendix \ref{sec:app:additional:valore} for details.

This analysis focuses on the introduction of the first retail Treasury bond \textit{Italia}, whose first issuance was announced on March 16, 2012, and took place from March 19 to March 22, 2012. The announcement detailed the bond's characteristics, including its four-year maturity, guaranteed minimum real coupon rate of $2.25\%$, and semi-annual coupon payments linked to the Italian Consumer Price Index [\textcite{mef2012b}]. The standard Treasury bond with ISIN $\mathit{IT0004805070}$, issued in March 2012 with a three-year maturity, serves as a useful reference for comparing the first issuance of retail Treasury bonds \textit{Italia}. This standard government bond offered a fixed coupon rate of $2.50\%$, providing semi-annual interest payments and predictable returns over its three-year maturity [\textcite{mef2012a}]. Unlike Treasury bonds \textit{Italia}, which are indexed to domestic inflation and designed specifically for retail investors, this bond was a conventional fixed-rate instrument targeting a broader investor base. Ex-ante, the first issuance of the retail Treasury bond \textit{Italia} appeared to be an attractive investment, particularly for retail investors concerned about inflation risk. Given its inflation protection, guaranteed minimum return, principal revaluation, and final bonus, it likely presented a more attractive investment compared to the standard Treasury bond, particularly in light of the positive inflation expectations in 2012.

The inaugural introduction of retail Treasury bonds \textit{Italia} received a positive response from retail investors, corresponding to about $7.3$ billion euros with $133,479$ contracts concluded [\textcite{mef2012c}]. By the end of 2012, Treasury bonds \textit{Italia} attracted a total of $27$ billion euros over three issuances [\textcite{mef2013}]. The contract size distribution confirms strong participation by small savers. Across all three issuances in 2012, more than $80\%$ of contracts were below the $50,000$--euro threshold typically used to distinguish retail from institutional investors. More than half of these contracts were even below $20,000$ euros, demonstrating the large participation of small savers.\footnote{
See \textcite{mef2012d}, \textcite{mef2012e}, and \textcite{mef2012f} for details on the placement of the first three retail Treasury bonds \textit{Italia} issuances. Since March 2012, cumulative take-up of \textit{Italia} amounts to $193.1$ billion euros across nineteen issuances as of 2025. The May 2025 issuance attracted over $8.79$ billion euros, with $65\%$ of subscriptions from individual investors and over $88\%$ of contracts below $50,000$ euros [\textcite{mef2025b}; \textcite{mef2025c}].}

In the empirical analysis, I examine the effects of introducing the first retail Treasury bonds \textit{Italia} on participation in the broader Treasury bond market, as the major impact can likely be observed immediately following the release of the new financial instrument.

\subsection{Retail Treasury bonds as a proxy for CBDC adoption} \label{proxy}
The empirical exercise uses the introduction of retail Treasury bonds in Italy as a proxy because evidence on CBDC adoption is not yet available. The proxy is not intended to mirror the policy objectives of CBDC. Rather, it is used to study a narrower question: how household heterogeneity shapes adoption when a new public retail instrument is introduced.

The motivation for the proxy is based on adoption mechanisms that may overlap between the two instruments. Retail Treasury bonds are government claims issued directly to households, and government issuance can matter for perceived safety and trust. Retail Treasury bonds are accessed through an institutional channel, and households can purchase them directly through home banking with an online trading feature. The decision to purchase them likely depends on how households evaluate the new instrument relative to alternative assets. Similarly, a retail CBDC would be offered and backed by the central bank, representing a safe public liability comparable to cash in terms of issuer risk. It would be accessed through a digital wallet at a financial institution, and the extent of adoption may depend on households’ assessment of its benefits and on how it compares to existing liquid assets. In both cases, assessment of the instruments’ functionality may vary across levels of financial literacy.

At the same time, important differences remain. Retail Treasury bonds are primarily investment products, and their issuance is driven by debt management and financing needs. A CBDC is primarily conceived as a means of payment, and future adoption patterns may be shaped by payment-specific forces such as merchant acceptance and network effects. These payment-side forces are central to the adoption of a new payment instrument and are not captured by the retail Treasury bond episode.

Within these limits, the proxy remains informative about heterogeneity in household adoption of a new public retail instrument, in particular across levels of financial literacy. The empirical evidence on differential participation following the retail Treasury bond introduction is used to discipline the theoretical model’s predictions for how CBDC demand may vary across households, rather than to draw conclusions about payment network dynamics.

\section{Survey data} \label{data}
To study the effects on market participation of the new financial instrument, I use data from the Bank of Italy's Survey on Household Income and Wealth (SHIW) on the incomes and savings of Italian households. The survey was conducted from 1960 to 1987 at yearly intervals on repeated cross-section observations. Starting from 1989, the survey runs every other year (except for 1998 and 2020) on repeated cross-section observation and panel households. The panel component of the sample consists of all households participating in at least two waves and an additional part randomly selected from those interviewed only in the previous edition.

The SHIW sample includes approximately $8,000$ households per wave, with around $4,000$ being panel households. The sample is designed to represent the Italian population, excluding people living in institutions or residing in the country illegally. In this section, I describe the data using the 2008, 2010, 2012, 2014, 2016, and 2020 waves.\footnote{
All descriptive statistics in the paper are computed using sampling weights. See Appendix \ref{sec:app:additional:weights} for details.
}

Each survey is conducted the year following the wave year. For example, the 2012 survey was administered in 2013. A typical question asked in the 2012 survey is \\

\noindent
\textit{Did you or a member of the household have any of the following} [financial assets] \textit{on 31 December 2012}: $\dots$.\\

Besides information on incomes and savings, the survey also provides information on household geographical and personal characteristics. Over the years, the survey has expanded to include wealth and various aspects of household economic behavior, such as financial assets and liabilities, and questions on financial knowledge.

\subsection{Financial assets} \label{datafina}
To investigate households' financial market participation, I construct five classes of financial assets as specified in Table \ref{assets}.\footnote{
The SHIW contains information on other financial assets excluded from the analysis as their individual magnitude is small, i.e., foreign securities, loans to cooperatives, private pensions, insurance policies for life and health, and against accidents.}
Real estate is excluded from the asset classes. Real estate is relatively important as a financial investment for many Italians, who often buy second houses as an investment. Considering the real estate equity value resulting from the difference between the value of the owned property and the debt incurred to buy it, real estate amounts to $80\%$ of the average financial asset investments in the data, and this fraction remains constant for most of the sample. However, in the survey, real estate includes the sum of all real estate, such as the primary residence, secondary houses (potentially rented out), commercial buildings, and land. While it is possible to distinguish between primary residences and other properties, the associated debt for purchasing these properties is only available in aggregate form. Therefore, this analysis focuses exclusively on purely financial assets. Specifically, as anticipated in Section \ref{itbonds}, the analysis focuses on household participation in the Treasury bond market, the highlighted category of government bonds in Table \ref{assets}.

\begin{figure}[!htbp]
\centering
\caption{Market participation of Italian households}
\begin{minipage}{\textwidth}
    \centering
    \makebox[\textwidth]{\includegraphics[width=0.8\textwidth]{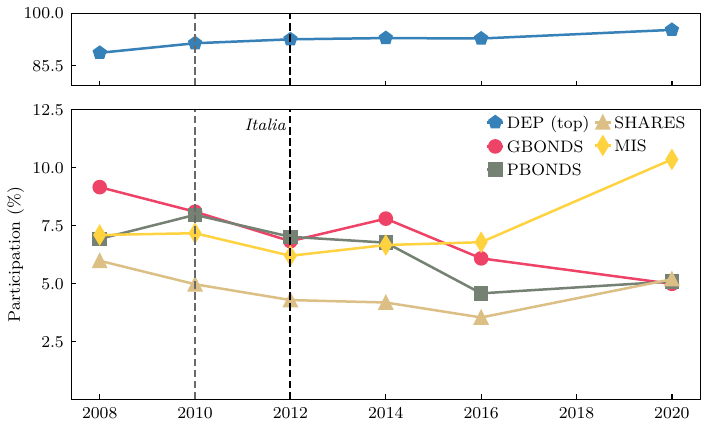}}
\end{minipage}
\vskip 0.5em
\begin{minipage}{\textwidth}
    \footnotesize
    \textit{Note}: This figure shows market participation across the sample of Italian households. Weighted average of the observations. \\
    \textit{Source}: SHIW and author calculations.
\end{minipage}
\label{fig:assetspart}
\end{figure}

Figure \ref{fig:assetspart} illustrates the average market participation of Italian households in the waves from 2008 to 2020. Households show a significant preference for traditional financial assets, with consistently high participation in deposits, which is reported in the top panel. Participation in government bonds shows a declining trend, falling from around $9\%$ in 2008 to approximately $6\%$ in 2020. Private bonds show a similar trend. Participation in more complex financial instruments, such as managed investment schemes and shares, remains low throughout most of the sample period, with a slight increase in the last wave.

\begin{table}[!htbp]
\centering
\caption{Financial asset classes}
\begin{threeparttable}
\begin{tabular}{llp{8cm}}
\toprule\toprule
\multicolumn{1}{c}{Class} & \multicolumn{1}{c}{Name} & \multicolumn{1}{c}{Individual assets in SHIW}\\
\midrule
Deposits & DEP & Bank current and saving accounts, CDs, repos, postal savings certificates \\
Government bonds & GBONDS & \textit{BOTs}, \textit{CCTs}, \textit{CTZs}, \textit{\textbf{BTPs}}, other Italian government securities \\
Private bonds & PBONDS & Privately issued bonds \\
Managed Investment Schemes & MIS & Managed funds shares, managed savings \\
Shares & SHARES & Shares\\
\bottomrule
\end{tabular}
\end{threeparttable}
\vskip 0.5em
\begin{minipage}{\textwidth}
    \footnotesize
    \textit{Note}: This table details individual assets in SHIW, grouped in classes. \textit{Buoni del Tesoro Ordinari (BOTs)} correspond to Treasury bills, \textit{Certificato di Credito del Tesoro (CCTs)} correspond to Treasury certificates, \textit{Certificato del Tesoro Zero-coupon (CTZs)} correspond to Zero-coupon bonds, and \textit{Buoni del Tesoro Poliennali (BTPs)} correspond to Treasury bonds with long-term maturity. \\
    The composition of classes may change over time based on whether questions about the ownership of specific assets have been included in the respective surveys. When occurring, the change is incremental, with new asset subclasses being added to the particular asset class and compositional changes only impacting previous measures. In general, these changes will not qualitatively affect the analysis results. \\
    \textit{Source}: SHIW.
\end{minipage}
\label{assets}
\end{table}

\begin{figure}[!htbp]
\centering
\caption{Government securities participation of Italian households}
\begin{minipage}{\textwidth}
    \centering
    \makebox[\textwidth]{\includegraphics[width=0.8\textwidth]{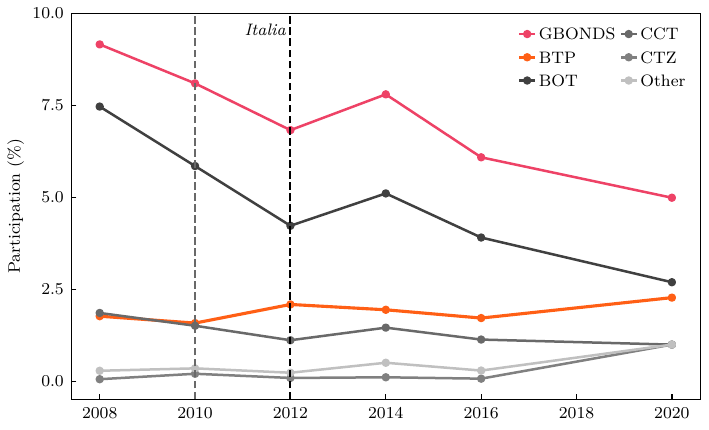}}
\end{minipage}
\vskip 0.5em
\begin{minipage}{\textwidth}
    \footnotesize
    \textit{Note}: This figure shows the breakdown of government securities participation across the sample of Italian households. Weighted average of the observations. \\
    \textit{Source}: SHIW and author calculations.
\end{minipage}
\label{fig:bondpart}
\end{figure}

Figure \ref{fig:bondpart} provides a detailed breakdown of participation in government securities. While Treasury bills (\textit{BOTs}) exhibit strong participation throughout the sample, their trend is sharply declining. In contrast, participation in Treasury bonds (\textit{BTPs}) remains low but relatively stable, with a slight upward trend. Participation in other government securities is lower across the entire sample period. The two vertical lines in Figure \ref{fig:bondpart} mark the 2010 wave preceding the policy introduction and the 2012 wave when the policy was implemented. Notably, participation in the Treasury bond market increases between these two waves, possibly reflecting the impact of the introduction of retail Treasury bonds \textit{Italia} in 2012.\footnote{
Recall that the 2012 survey was administered in 2013. Therefore, participation in 2012 reflects the end of the year, influenced by the introduction of the first retail Treasury bonds \textit{Italia} in March 2012.
}

\subsection{Financial literacy} \label{datafl}
\textcite{Hastings2013} define financial literacy as ``the ability to use knowledge and skills to manage one's financial resources effectively for lifetime financial security'', while other definitions refer to knowledge of financial concepts or products. Financial literacy measures in the literature encompass various methodologies and indicators to assess individuals' comprehension of fundamental financial concepts. One of the most common ways to identify financially literate respondents is based on their accurate responses to the ``Big Three'' questions [see, e.g., \textcite{Gallo2023}], as defined by \textcite{Lusardi2014}, which capture the basic financial concepts of interest, inflation, and risk diversification. These questions assess essential financial knowledge by testing individuals' understanding of how interest compounds over time, the effects of inflation on purchasing power, and the importance of diversifying investments to mitigate risk. Together, they provide a concise measure of financial literacy that is widely used in research to gauge people's capability to make sound financial decisions.

Other measures of financial literacy in the literature include the number of correct responses to the ``Big Five'' questions proposed by \textcite{Hastings2013}, emphasizing a broader scope of financial knowledge [see, e.g., \textcite{Angrisani2023}]. \textcite{Dalessio2020} adopt the OECD methodology and construct the financial literacy score from assessments of knowledge, behavior, and attitude. To evaluate the reliability of each component, the authors use the Cronbach coefficient, which assesses internal coherence of the different items that make up each component: The higher the correlation among items and the greater their number, the higher the coefficient. Notably, they find that the most reliable indicator is financial knowledge, which captures the basic financial concepts reflected in the ``Big Three'' questions.

Since 2006, the SHIW has collected information on household financial knowledge (except for the 2012 and 2014 waves).\footnote{
See Appendix \ref{sec:app:additional:shiwq} for the exact formulation of the questions.} 
The questions vary in number and content over time. However, the ``Big Three'' questions are consistently present across the waves. The question related to interest rates appears in the 2006, 2016, and 2020 waves; the question regarding risk diversification is asked in the 2008, 2010, 2016, and 2020 waves; and the question about inflation is present in all the waves considered in the analysis. Additionally, the survey includes a question assessing the financial knowledge of mortgages in the 2006, 2008, and 2010 waves. In the 2006 wave, financial literacy questions were asked to a subset of families, therefore I do not include this wave in the statistics.

\begin{figure}[!htbp]
\centering
\caption{Correct responses to financial literacy question}
\begin{minipage}{\textwidth}
    \centering
    \makebox[\textwidth]{\includegraphics[width=0.8\textwidth]{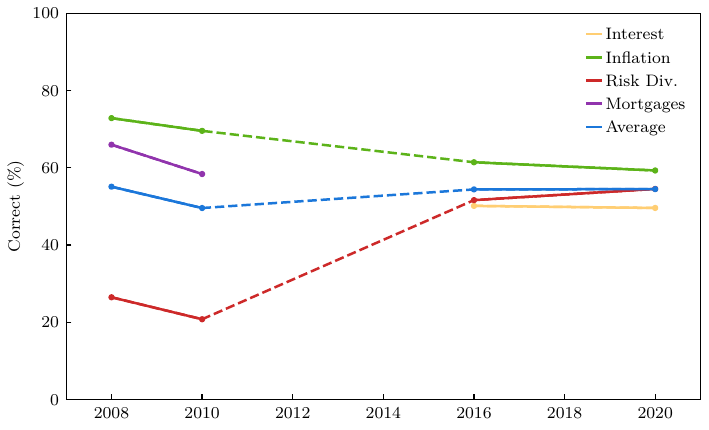}}
\end{minipage}
\vskip 0.5em
\begin{minipage}{\textwidth}
    \footnotesize
    \textit{Note}: This figure shows the aggregate percentage of correct responses to financial literacy questions. Incorrect responses include ``Don't know'' and ``No answer''. The average refers to the average number of correct answers. \\
    \textit{Source}: SHIW and author calculations.
\end{minipage}
\label{fig:correct}
\end{figure}

Figure \ref{fig:correct} shows the aggregate percentage of correct responses to financial literacy questions and the average number of correct answers across the cross-section of households in different waves of the SHIW. Compared to responses to other questions, the inflation question consistently has higher correct responses, particularly in 2008 and 2010, though it shows a decline in 2020. The risk diversification question has lower correct response rates, peaking in 2020. The interest rate question, asked in the last two waves, shows correct response rates of around $50\%$. The mortgage question, included in the earlier waves, shows relatively high correct response rates, especially in 2008. The average number of correct answers indicates a stable level of financial literacy over time, with slight fluctuations.\footnote{See Table \ref{tab:correct} in Appendix \ref{sec:app:additional} for the percentage numbers of correct responses.}

I define the financial literacy score as the number of correct answers to the financial literacy questions. Each wave includes a different combination of three questions, so the financial literacy score can have four levels: $0$ if the household does not answer any question correctly, $1$ if there is one correct answer, $2$ if there are two correct answers, and $3$ if all answers are correct.\footnote{
See Appendix \ref{sec:app:additional:sumstat} for descriptive statistics for the financial literacy score by demographics, across Italian regions, and wealth distribution.}

\begin{figure}[!htbp]
\centering
\caption{Market participation by literacy score level}
\begin{minipage}{\textwidth}
    \centering
    \makebox[\textwidth]{\includegraphics[width=0.8\textwidth]{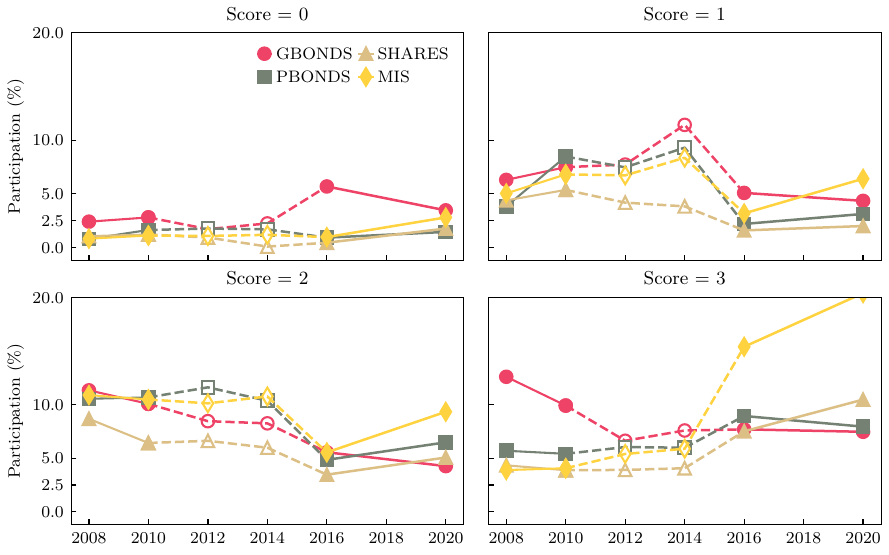}}
\end{minipage}
\vskip 0.5em
\begin{minipage}{\textwidth}
    \footnotesize
    \textit{Note}: This figure shows market participation across the sample of Italian households, distinguishing between financial literacy score levels. When the information on household literacy scores is missing for a specific year, the last data available are used. Weighted average of the observations. Deposits are omitted. \\
    \textit{Source}: SHIW and author calculations.
\end{minipage}
\label{fig:assetpartlev}
\end{figure}

\begin{figure}[!htbp]
\centering
\caption{Government securities participation by literacy score level}
\begin{minipage}{\textwidth}
    \centering
    \makebox[\textwidth]{\includegraphics[width=0.8\textwidth]{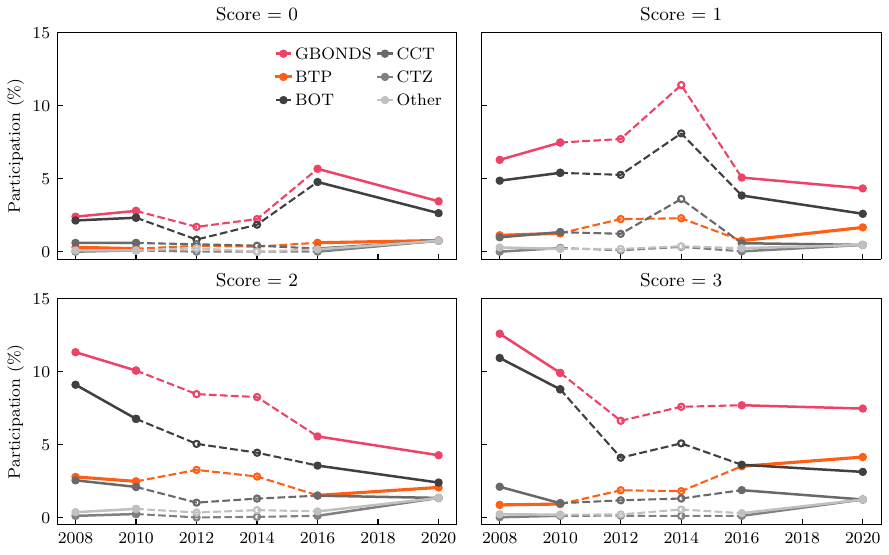}}
\end{minipage}
\vskip 0.5em
\begin{minipage}{\textwidth}
    \footnotesize
    \textit{Note}: This figure shows the participation in government securities markets across the sample of Italian households, distinguishing between financial literacy score levels. When the information on household literacy scores is missing for a specific year, the last data available are used. Weighted average of the observations. \\
    \textit{Source}: SHIW and author calculations.
\end{minipage}
\label{fig:bondpartlev}
\end{figure}

Similar to Figures \ref{fig:assetspart} and \ref{fig:bondpart} in Section \ref{datafina}, Figures \ref{fig:assetpartlev} and \ref{fig:bondpartlev} report overall market participation and participation in government securities, respectively, distinguishing between financial literacy score levels. Figure \ref{fig:assetpartlev} suggests that households with higher financial literacy scores tend to hold more diversified portfolios. However, in the 2012 wave, households with the highest score level 3 appear less diversified than those with levels 1 and 2. A similar pattern emerges in Figure \ref{fig:bondpartlev}, which shows the breakdown by government securities. Focusing on the Treasury bond (\textit{BTP}) market, participation differs across financial literacy levels. This evidence underscores the role of financial literacy in influencing investment decisions.

\begin{figure}[!htbp]
\centering
\caption{Percentage of households by financial literacy score level}
\begin{minipage}{\textwidth}
    \centering
    \makebox[\textwidth]{\includegraphics[width=0.8\textwidth]{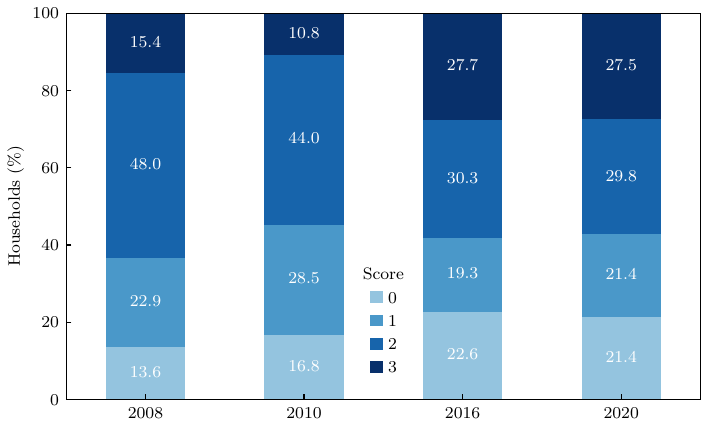}}
\end{minipage}
\vskip 0.5em
\begin{minipage}{\textwidth}
    \footnotesize
    \textit{Note}: This figure shows the percentage of households by financial literacy score level. \\
    \textit{Source}: SHIW and author calculations.
\end{minipage}
\label{fig:finlitlev}
\end{figure}

Finally, Figure \ref{fig:finlitlev} shows the percentage of households by financial literacy score level in four waves since 2008, excluding 2012 and 2014 when the financial literacy questions were not included. Over time, the distribution becomes more evenly spread across financial literacy levels.

\section{Retail Treasury bonds and market participation} \label{empexercise}
This section evaluates the effect of the introduction of retail Treasury bonds on household participation in the broader Treasury bond market. Previous studies have shown that financial literacy influences household financial behavior [see, e.g., \textcite{Arrondel2016}, \textcite{Vanrooij2011}, and \textcite{ChenDai2023}]. The empirical model estimates the impact of the policy on participation across households with different levels of financial literacy, thereby quantifying the differential policy-financial literacy effect.

The policy used is the introduction of retail Treasury bonds \textit{Italia} in March 2012. These are the first retail Treasury bonds issued, and likely, the major effects of the retail government policy on market participation can be seen right after their introduction. To purchase retail Treasury bonds, an individual must have an investment account with a bank or financial institution. Before opening an investment account, banks in Italy are required to assess the individual's financial knowledge and experience. This is typically done through a profiling questionnaire as part of the Markets in Financial Instruments Directive regulations. This questionnaire aims to evaluate the individual's understanding of financial products and ensure that the bank offers suitable investments based on the client's risk tolerance and knowledge. However, the questions in the questionnaire are not directly comparable to the ``Big Three'' questions commonly used to assess financial literacy, as they focus less on basic financial concepts and more on practical aspects related to investment behavior.

The financial literacy score defined in Section \ref{datafl} can take four levels, each identifying a distinct household type: $0$ for financially illiterate, $1$ for low-financially literate, $2$ for high-financially literate, and $3$ for financially expert. I conduct a short-run analysis using data from the 2010 and 2012 SHIW waves, where the household's initial level of financial literacy is the one in 2010, and the policy is introduced in March 2012 (recall that the 2012 survey was run in 2013). 

\begin{figure}[!htbp]
\centering
\caption{Timeline of the short-run analysis}
\resizebox{0.8\textwidth}{!}{%
    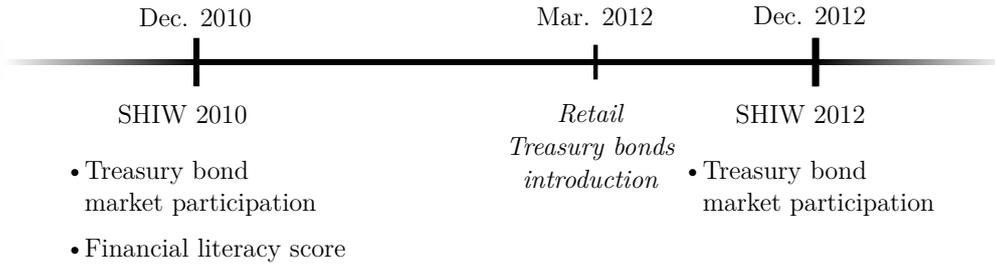%
}
\label{fig:timeline}
\end{figure}

Figure \ref{fig:timeline} shows the timeline of the policy introduction and the data. The panel dimension of the data allows me to follow the investment behavior of the same household over time. A traditional empirical challenge is reverse causality: households with higher wealth may have more opportunities and incentives to acquire financial skills and, in turn, increase their wealth [\textcite{Lusardi2017}]. Exploiting the longitudinal dimension of the dataset, I circumvent the problem of reverse causality: household asset-holding decisions in 2012 cannot affect their level of financial knowledge in 2010.

\subsection{Baseline regression}
I construct the empirical model using the Generalized Linear Model (GLM) framework, which extends linear regression to binary outcomes. Instead, using a Linear Probability Model (LPM) can produce biased estimates and inaccurate standard errors with binary outcomes and skewed distributions (such as the low initial share of Treasury bond investors), as the LPM's assumptions of homoscedasticity and normally distributed residuals do not hold for binary data.

The relationship between predictors and the outcome mean is modeled via a logistic link function, $g$. Let $y_{it}$ denote the observed Treasury bond market participation of household $i$ at time $t$, where $1$ indicates participation and $0$ otherwise, and let $P_{it} = \mathbb{E}[y_{it}]$ represent the probability of participation. The probability $P_{it}$ is determined by the latent variable $z_{it}$, which captures the propensity of the household to participate, through the logit link function:
\begin{align}
    z_{it} = g(P_{it}) \equiv \ln \left( \frac{P_{it}}{1-P_{it}} \right) \label{linkf}.
\end{align}
The propensity $z_{it}$ is modeled as:
\begin{align}
    z_{it} = \alpha + \beta ({D}^{12}_t \times {LS}^{10}_i) + \sum_{\ell=1}^{3} \gamma_\ell {LS}^{10}_{i \ell} + \boldsymbol{\eta}^\intercal {\boldsymbol{X}}_{it} + \epsilon_{it}, \quad {t = 2010, 2012}, \label{baseline}
\end{align}
where the coefficient of interest, $\beta$, captures the interaction between the policy dummy $D_t^{12}$, which equals $1$ in 2012, and the initial household financial literacy score level, $LS^{10}_{i}$. This score can take four values corresponding to the same number of household types: $0$ if financially illiterate, $1$ if low-financially literate, $2$ if high-financially literate, and $3$ if financially expert.

The natural propensity to participate for financial literacy level $0$ is captured by $\alpha$, and group-fixed effects for each financial literacy level $\ell$ are accounted for by the $\gamma_\ell$ coefficients. Additionally, the coefficient vector $\boldsymbol{\eta}$ captures the effects of the vector of controls $\boldsymbol{X}_{it}$. Following the household finance and financial literacy literature, the set of controls includes household demographic characteristics (age, age squared, a dummy for being married, and a dummy for being female) to capture life-cycle and household structure [\textcite{Angrisani2023}, \textcite{Calvet2009}, \textcite{Vanrooij2011}]. Education is excluded because it correlates with financial literacy and would absorb part of the variation used to study heterogeneity by literacy. Wealth is also excluded because it may lie on the pathway from financial literacy to market participation through long-run saving and portfolio choices, and, in this setting, may also be jointly determined with security holdings. Additionally, an indicator of high risk aversion measured in 2010 is included to account for preference heterogeneity [\textcite{Guiso2009}, \textcite{Hsiao2018}, \textcite{Vanrooij2011}].

The estimated coefficient of interest, $\hat{\beta}$, can be interpreted through the odds ratio:\footnote{
The odds of an outcome occurring is the ratio of successes to failures. See Appendix \ref{sec:app:emp:odds} for the intuition for interpreting the estimated coefficient of interest.
}
\begin{align}
    \frac{P_1}{1-P_1} = \frac{P_0}{1-P_0} \times \exp(\hat{\beta}).
\end{align}
For the high-financially literate household (i.e., when $D^{12} \times {LS}^{10} = 2$), the odds ratio can be evaluated as:
\begin{align}
    \frac{P_2}{1-P_2} = \frac{P_{1}}{1-P_{1}} \times \exp(\hat{\beta}) = \frac{P_{0}}{1-P_{0}} \times \exp(2\hat{\beta}),
\end{align}
where $\exp(\hat{\beta})$ represents the multiplicative change in the odds (i.e., the odds ratio) for each one-unit increase in the policy-financial literacy interaction $D^{12} \times {LS}^{10}$.

The intuition is that, after introducing retail Treasury bonds, each higher level of financial literacy score changes the odds of participating in the broader Treasury bond market by $\exp(\hat{\beta})$ compared to the level below. Specifically, if $\exp(\hat{\beta})$ is greater than $1$, it implies an increase in the odds, whereas a value less than $1$ indicates a decrease in the odds.

\subsection{Financial literacy score dummies regression}
Alternatively, to explore potential non-linear relationships across financial literacy levels, I model the household $i$ propensity to participate, $z_{it}$, as
\begin{align}
    z_{it} = \alpha + \sum_{\ell=1}^{3} \beta_{\ell} ({D}^{12}_t \times {LS}^{10}_{i \ell}) + \sum_{\ell=1}^{3} \gamma_\ell {LS}^{10}_{i \ell} + \boldsymbol{\eta}^\intercal {\boldsymbol{X}}_{it} + \epsilon_{it}, \quad {t = 2010, 2012}, \label{level}
\end{align}
with $LS^{10}_{i\ell}$ being household financial literacy score dummies for each score level $\ell$, where the illiterate (i.e., score level $0$) are the baseline.

The estimated coefficients of interest, $\hat{\beta_{\ell}}$, can be interpreted through the odds ratio:
\begin{align}
    \frac{P_\ell}{1-P_\ell} = \frac{P_0}{1-P_0} \times \exp(\hat{\beta}_\ell), \quad \ell = 1,2,3.
\end{align}
For instance, for the high-financially literate group (i.e., $\ell = 2$):
\begin{align}
    \frac{P_2}{1-P_2} = \frac{P_0}{1-P_0} \times \exp(\hat{\beta}_2).
\end{align}

The intuition is analogous to before: Following the introduction of retail Treasury bonds, the odds of a high-financially literate household participating in the Treasury bond market are $\exp(\hat{\beta}_2)$ times higher compared to the financial illiterate household baseline.

As previously mentioned, the main aim of the empirical investigation is to explore if and how financial literacy influenced households' participation in the Treasury bond market following the introduction of the new instrument. If so, it is interesting to investigate the potential household portfolio reallocation. Therefore, participation in government securities other than Treasury bonds and stock holdings are considered alternative output variables.\footnote{
I define stock holdings as holding either shares or managed investment schemes, as specified in Table \ref{assets}, following \textcite{Calvet2009}, to capture direct and indirect equity market participation.
}

\subsection{Results}\label{empresult}
Table \ref{logitoods} reports the estimated differential changes in the odds of participating. Recall that the linear baseline specification (\ref{baseline}) uses financial literacy score levels for illiterate, low-literate, high-literate, and expert. In the SHIW data, the percentage of households participating in the Treasury bond market at the end of 2010 was $1.58$\%, and at the end of 2012 it was $2.08\%$. After the introduction of retail Treasury bonds in March 2012, each higher level of financial literacy score increases the odds of participating in the Treasury bond market by $15\%$ relative to the next-lower literacy level, as reported in column (1).\footnote{
Table \ref{logit} in Appendix \ref{sec:app:emp} reports the logit regression estimation results. Recall that the odds of participating change by, e.g., $\exp(\hat{\beta}) -1 = \exp(0.14) -1 = 15\%$.} 
Given the low baseline participation rate, this estimate can be interpreted as a meaningful relative shift on the participation margin. When considering alternative market participation dependent variables, column (3) shows that each level increase in the financial literacy score decreases the odds of participating in the market for other government securities by $19$\%. The odds of stock-holding decrease by $5\%$ for each additional level of financial literacy score, as reported in column (5).

When considering the breakdown by financial literacy score level in the data, Treasury bond market participation rates at the end of 2010 were $0.21\%$, $1.26\%$, $2.47\%$, and $0.92\%$ for financially illiterate, low-financially literate, high-financially literate, and financially expert, respectively.\footnote{
Note that information on household financial literacy is missing for 2012 in the SHIW data.
} The results from the non-linear specification (\ref{level}) suggest that, after the new instrument introduction, as literacy increases, the odds of participating in the Treasury bonds market also increase compared to the baseline for the financially illiterate. However, the rate of increase is not monotone across financial literacy levels, as indicated by the coefficients of low and high literacy in column (2) in Table \ref{logitoods}: while for the low-financially literate, the odds of participating in the Treasury bond market increase by $80\%$ compared to the financially illiterate, the increase is only $25\%$ for the high-financially literate ($57\%$ for the financially expert). Similarly, the decrease in the odds of stock holdings, compared to the baseline of illiterate, is more pronounced for the low-literate, as illustrated in column (6). In contrast, column (4) shows that the decrease is greater for the highly literate and expert when considering participation in the market for other government securities.

\begin{table}[!htbp]
\centering
\caption{Changes in the odds of participation}
\resizebox{\textwidth}{!}{\begin{threeparttable}
\begin{tabular}{lSSSSSS}
\toprule\toprule
\multirow{2}{*}{\begin{tabular}[c]{@{}l@{}}Dep. Var.: \\ Market Participation\end{tabular}} 
& \multicolumn{2}{c}{Treasury bond} & \multicolumn{2}{c}{Other Govt. Bond} & \multicolumn{2}{c}{Stocks}\\
\cmidrule(lr){2-3} \cmidrule(lr){4-5} \cmidrule(lr){6-7}
& \multicolumn{1}{c}{(1)} & \multicolumn{1}{c}{(2)} & \multicolumn{1}{c}{(3)} & \multicolumn{1}{c}{(4)} & \multicolumn{1}{c}{(5)} & \multicolumn{1}{c}{(6)} \\
\midrule
$D^{12} \times {LS}^{10}$ & 0.15\textsuperscript{***} & & -0.19\textsuperscript{***} & & -0.05\textsuperscript{**} & \\
$D^{12} \times {LS}_1^{10}$ & & 0.80\textsuperscript{**} & & -0.24\textsuperscript{**} & & -0.21\textsuperscript{***} \\
$D^{12} \times {LS}_2^{10}$ & & 0.25\textsuperscript{*} & & -0.34\textsuperscript{***} & & -0.12\textsuperscript{**} \\
$D^{12} \times {LS}_3^{10}$ & & 0.57\textsuperscript{*} & & -0.47\textsuperscript{***} & & 0.05\text{} \\
\midrule
Test of $H_0: \hat{\beta}_1 \leq \hat{\beta}_2$ & \multicolumn{2}{c}{$\hat{\beta}_1 = 0.80$, $\hat{\beta}_2 = 0.25$, $\text{p-value} = 0.09$} & & & & \\
\bottomrule
\end{tabular}
\end{threeparttable}
}
\vskip 0.5em
\begin{minipage}{\textwidth}
    \footnotesize
    \textit{Note}: This table reports changes in the odds of participating. The changes are computed as \\
    \noindent\makebox[\textwidth]{\centering $\exp(\hat{b})-1, \quad b \in \{ \beta, \beta_{\ell} \}$,}\\
    where the coefficients $\beta$ and $\beta_\ell$ are estimated using the Generalized Estimating Equations (GEE), an extension of GLM for panel data. The sample counts a panel of 4,611 households over a two-year period. Literacy level group-fixed effects and time-fixed effects are included. All regressions include a vector of baseline demographic controls and a dummy for high risk aversion in 2010.\\
    Columns (1), (3), and (5) refer to specification (\ref{baseline}), while columns (2), (4), and (6) refer to specification (\ref{level}). \\
    In specification (\ref{level}), the baseline financial literacy level is 0. * $p < 0.1$; ** $p < 0.05$; *** $p < 0.01$.\\
    Test of $H_0: \hat{\beta}_1 \leq \hat{\beta}_2$ for low- vs. high-literate households in Column (2) results in one-sided p-value = $0.09$.
\end{minipage}
\label{logitoods}
\end{table}

I formally test whether the effect on Treasury bond market participation for low-financially literate households exceeds that for the high-literacy group (i.e., compare the coefficients in the second and third row in column (2) in Table \ref{logitoods}). Using a one-sided Wald test, I reject the null at the $10\%$ significance level, indicating that the estimated coefficient for low-literacy households is statistically larger than that for high-literacy households. See Appendix \ref{sec:app:emp:inference} for details.

To summarize, from the baseline linear regression (\ref{baseline}), I find that financial literacy positively influenced the propensity to participate in the broader Treasury bond market after the introduction of retail Treasury bonds. However, the effect is non-monotone across financial literacy levels, as shown in the results from the financial literacy score dummies specification (\ref{level}). From column (2) in Table \ref{logitoods}, households with some but low financial literacy are more likely to participate compared to the other household groups. One interpretation for this result is a fixed cost of entry mechanism. Baseline participation in the Treasury bond market is low in the SHIW data, suggesting that even small information costs can keep households out. The launch of the retail instrument was accompanied by a dedicated institutional communication campaign, which plausibly increased awareness and reduced informational barriers for retail investors [\textcite{mef2011b}]. The new instrument may also have reduced participation frictions by offering a simpler and more accessible retail channel, which is especially relevant for households that have some ability to engage with financial products but still face higher frictions than high-financially literate households. Financially illiterate households may remain constrained by limited awareness or procedural barriers, while highly literate and expert households may be more inclined to allocate to other, more attractive investment opportunities. Lastly, when using alternative dependent variables, results suggest a portfolio shift from other government bonds and stocks toward Treasury bonds. Interestingly, from column (6) in Table \ref{logitoods}, following the new policy, high-financially literate households are less likely to decrease participation in the stock market compared to the other household groups.

Table \ref{logit_robustness} in Appendix \ref{sec:app:emp:robustness} reports the logit regression estimation results using the covariates selected by a post-double-selection procedure with the SCAD penalty. The qualitative findings remain unchanged. Under the linear specification, following the introduction of retail Treasury bonds, the propensity to participate in the Treasury bond market increases with financial literacy, while that in the other government bonds and the stock market declines. Under the non-linear specification, the non-monotonic heterogeneity pattern across literacy levels is preserved.

\section{A simple two-agent endowment economy}\label{theory}
This section explores the potential implications of financial literacy for CBDC adoption. While CBDC could be accessible to everyone, its adoption depends on several factors, including household financial literacy.

Given the lack of empirical data on CBDC usage, the introduction of retail Treasury bonds serves as a proxy for studying CBDC adoption. Findings from the retail Treasury bonds exercise in Section \ref{empexercise} suggest a differential financial behavior depending on financial literacy, with households in the low-literate group being more likely to participate following the introduction of the new instrument compared to the financially illiterate and high-financially literate household groups.

Building on these empirical insights, I develop a simple illustrative model to explore how CBDC demand may vary by financial literacy. In the theoretical model, I abstract from modeling the financially illiterate group, as I want to disentangle the potential different CBDC adoption conditional on the financial literacy level, high or low. For simplicity, I also abstract from modeling the financially expert group as separate from the high-literacy group, as their financial behavior does not differ significantly in the data.

\begin{figure}[!htbp]
\centering
\caption{Portfolio composition across Italian households}
\begin{minipage}{\textwidth}
    \centering
    \makebox[\textwidth]{\includegraphics[width=0.8\textwidth]{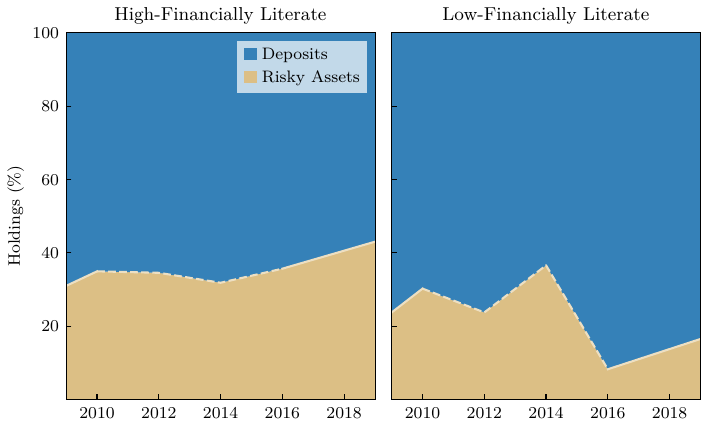}}
\end{minipage}
\vskip 0.5em
\begin{minipage}{\textwidth}
    \footnotesize
    \textit{Note}: This figure shows the overall portfolio allocation across a sample of Italian households, distinguishing between high and low literacy groups. High-financially literate corresponds to households with literacy score levels 2 and 3, and low-financially literate to households with literacy score level 1. Risky assets include shares and managed investment schemes [\textcite{Calvet2009}]. Holdings are normalized such that deposits and risky assets together account for the entire portfolio. When the information on household literacy scores is missing for a specific year, the last data available are used. \\
    \textit{Source}: SHIW and author calculations.
\end{minipage}
\label{fig:shiwcross}
\end{figure}

The economy is populated by two types of agents differentiated by their financial literacy: high-financially literate (HFL) and low-financially literate (LFL). The HFL agent can invest in risky assets and risk-free deposits, whereas the LFL agent is limited to deposits. This distinction between agents' investment opportunities is motivated by the SHIW data. Figure \ref{fig:shiwcross} shows the portfolio composition across the average sample of Italian households, distinguishing between high and low financial literacy groups. On average, HFL households tend to hold more risky assets than LFL households. Additionally, there is a downward trend in risky asset holdings among LFL households. The literature has shown that information costs help explain why some individuals choose not to hold risky assets [\textcite{Vissing2003}]. The higher costs of processing financial information make LFL households more reluctant to invest in the risky asset market. To incorporate this aspect, in the model I consider the limit scenario where the LFL agent does not hold any risky assets.\footnote{
This assumption should be interpreted as an extreme benchmark capturing limited access to risky assets among the LFL agent, rather than as a literal description of zero participation. The mechanism in the model relies on an asymmetry in portfolio opportunities across literacy groups. Allowing the LFL agent to hold a small risky position or introducing explicit participation costs would be a natural extension for future work.
}

Each agent lives for two periods and receives an endowment $y$ in the first period. In the second period, agents receive income $\epsilon$ such that
\begin{align} \label{incomeshock}
  \epsilon =
    \begin{cases}
      {s}_{\mathrm{max}} & \text{with probability $p^\epsilon$},\\
      {s}_{\mathrm{min}} & \text{with probability $1-p^\epsilon$}.
    \end{cases}       
\end{align}
In the portfolio choice analysis, I consider cases with negative minimum income ${s}_{\mathrm{min}}$, which represents a loss of the income source and unexpected cost in the second part of the agents' working life.

Agents maximize their discounted expected lifetime utility, $\mathbb{E}[\mathcal{U}]$, subject to budget constraints for each period. Each agent values liquidity services derived from holding financial assets and has a concave, additively separable utility function in consumption.

\subsection{Pre-CBDC economy}
In the pre-CBDC economy, both agents can access risk-free deposits. Assuming log-utility:
\begin{align}
    \mathbb{E}[\mathcal{U}^j] = \ln(c_1^j) +\beta \mathbb{E} \ln(c_2^j) + \gamma \ln \big ( z(d^j) \big ), \quad {j = h, l},
\end{align}
where $j$ denotes the agent type $h$ or $l$ for the high- and low-financially literate agent, respectively; $c_1^j$ and $c_2^j$ are consumption in the first and second period; and $\beta \in (0,1)$ is the discount factor. The last term captures the liquidity benefits of holding deposits, with $\gamma \geq 0$ the liquidity preference parameter, and liquidity services $z$ a function of deposits $d^j$:
\begin{align}
    z(d^j) = d^j, \quad{j= h, l}.
\end{align}

The HFL agent can invest in risk-free deposits with return $R^d>1$ and risky assets $a$ with stochastic return $R^a$ such that
\begin{align} \label{ra}
  R^a =
    \begin{cases}
      \bar{R}^a>R^d & \text{with probability $p^a$},\\
      \ubar{R}^a<1 & \text{with probability $1-p^a$},
    \end{cases}       
\end{align}
where $\bar{R}^a$ and $\ubar{R}^a$ are the maximum and minimum return, respectively.

The HFL agent's budget constraints in the first and second periods are:
\begin{align}
    c_1^h+a+d^h =& y,\\
    c_2^h =& R^a a+R^d d^h+\epsilon.
\end{align}
The HFL agent maximizes the discounted expected lifetime utility
\begin{align}
    \mathbb{E}[\mathcal{U}^h] =& \ln(y-a-d^h)+\beta \bigg \{ p^a \left[p^\epsilon \ln(\bar{R}^a a+R^d d^h +{s}_{\mathrm{max}}) \right. \nonumber \\
    & \hspace{5cm} \left. + (1-p^\epsilon) \ln(\bar{R}^a a + R^d d^h + {s}_{\mathrm{min}}) \right] \nonumber \\
    & \hspace{3.5cm} + (1-p^a) \left[p^\epsilon \ln(\ubar{R}^a a + R^d d^h + {s}_{\mathrm{max}}) \right. \nonumber \\
    & \hspace{5cm} \left. +(1-p^\epsilon) \ln(\ubar{R}^a a + R^d d^h + {s}_{\mathrm{min}})\right] \bigg \} \nonumber \\
    & \hspace{3.5cm} + \gamma \ln(d^h),
\end{align}
by choosing optimal quantities of assets $a$ and $d^h$. 

Note that there is an implicit constraint due to the domain of the logarithmic utility function such that the problem is well defined only when
\begin{equation}\label{consth}
    \ubar{R}^a a + R^d d^h + {s}_{\mathrm{min}} > 0 \,.
\end{equation}
This mathematical constraint reflects the need of the agent to safeguard against the worst-case scenario of a negative return on the risky asset investment combined with a low minimum income in the second period, regardless of the probability of this event.

The LFL agent has access only to deposits and faces the following budget constraints in the first and second periods, respectively:
\begin{align}
    c_1^l+d^l =& y,\\
    c_2^l =& R^d d^l+\epsilon.
\end{align}
The LFL agent maximizes the discounted expected lifetime utility
\begin{align}
    \mathbb{E}[\mathcal{U}^l] = \ln(y-d^l)+\beta \left[ p^\epsilon \ln(R^d d^l +{s}_{\mathrm{max}}) + (1-p^\epsilon) \ln(R^d d^l + {s}_{\mathrm{min}}) \right] + \gamma \ln(d^l),
\end{align}
by choosing the optimal quantity of deposits $d^l$.

In this case, the implicit constraint due to the domain of the logarithmic utility function is
\begin{equation}\label{constl}
    R^d d^l + {s}_{\mathrm{min}} > 0 \,.
\end{equation}

\subsection{CBDC economy}
In the CBDC economy, the government introduces a CBDC accessible to both types of households, interest-bearing, and offering benefits relative to deposits. Importantly, the relation between financial instrument returns is as follows:
\begin{align}
    \mathbb{E}[R^a]  > R^d \geq R^m > 1,
\end{align}
such that the expected return on risky assets, $\mathbb{E}[R^a]$, is greater than the return on risk-free deposits, $R^d$, which in turn is assumed to be higher than or equal to the return on CBDC, $R^m$.

Agents' discounted expected lifetime utility is now assumed to follow:
\begin{align*}
    \mathbb{E}[\mathcal{U}^j] = \ln(c_1^j) +\beta \mathbb{E} \ln(c_2^j) + \gamma \ln \big ( z(d^j,m^j) \big ), \quad{j = h, l},
\end{align*}
where liquidity services $z$ are a function of deposits $d^j$ and CBDC $m^j$, as defined by:
\begin{align}
    z(d^j,m^j) = \left[(d^j)^{1-\sigma} + \lambda (m^j)^{1-\sigma}\right]^\frac{1}{1-\sigma} \label{liqserv}.
\end{align}
In equation (\ref{liqserv}), CBDC and deposits are imperfect substitutes, with $\sigma \geq 0$ representing the inverse elasticity of substitution between the two liquid assets.\footnote{
Several studies in the literature consider imperfect substitutability between CBDC and deposits [see, e.g., \textcite{Agur2022}, \textcite{Bacchetta2025}, \textcite{Barrdear2022}, \textcite{Burlon2022}, \textcite{Kumhof2021}, and \textcite{Niepelt2024}]. I use a constant elasticity of substitution (CES) aggregator for liquidity services as a tractable way to model substitution between CBDC and deposits with a single parameter.
} The parameter $\lambda \geq 0$ indexes the agent's preference for CBDC relative to deposits, reflecting factors such as liquidity benefit, privacy, or convenience of use.

With CBDC, the HFL agent's budget constraints are:
\begin{align}
    c_1^h+a+d^h+m^h =& y,\\
    c_2^h =& R^a a+R^d d^h+R^m m^h+\epsilon.
\end{align}
The HFL agent maximizes the discounted expected lifetime utility
\begin{align}
    \mathbb{E}[\mathcal{U}^h] =& \ln(y-a-d^h-m^h)+\beta \bigg \{ p^a \left[p^\epsilon \ln(\bar{R}^a a+R^d d^h+R^m m^h+{s}_{\mathrm{max}}) \right. \nonumber \\
    & \hspace{5cm} \left. + (1-p^\epsilon) \ln(\bar{R}^a a + R^d d^h +R^h m^h + {s}_{\mathrm{min}}) \right]  \nonumber \\
    & \hspace{3.5cm} + (1-p^a) \left[p^\epsilon \ln(\ubar{R}^a a + R^d d^h+R^m m^h + {s}_{\mathrm{max}}) \right. \nonumber \\
    & \hspace{5cm} \left. +(1-p^\epsilon) \ln(\ubar{R}^a a + R^d d^h+R^m m^h + {s}_{\mathrm{min}}) \right] \bigg \} \nonumber \\
    & \hspace{3.5cm} + \gamma \ln \big ( z(d^h,m^h) \big ),    
\end{align}
by choosing optimal quantities of assets $a$, $d^h$, and $m^h$.

With CBDC, the implicit constraint due to the domain of the logarithmic utility function reads
\begin{equation}\label{consthcbdc}
    \ubar{R}^a a + R^d d^h + R^m m^h + {s}_{\mathrm{min}} > 0 \,.
\end{equation}

Now the LFL agent can access a new financial instrument. Their budget constraints are:
\begin{align}
    c_1^l+d^l+m^l =& y,\\
    c_2^l =& R^d d^l+R^m m^l+\epsilon.
\end{align}
The LFL agent maximizes the discounted expected lifetime utility
\begin{align}
    \mathbb{E}[\mathcal{U}^l] =& \ln(y-d^l-m^l)+\beta \left[ p^\epsilon \ln(R^d d^l+R^m m^l +{s}_{\mathrm{max}})\right. \nonumber \\
    & \hspace{3.5cm} \left. + (1-p^\epsilon) \ln(R^d d^l +R^m m^l + {s}_{\mathrm{min}}) \right] \nonumber \\
    & \hspace{2.5cm} + \gamma \ln \big ( z(d^l,m^l) \big ),
\end{align}
by choosing optimal quantities of $d^l$ and $m^l$.

The implicit constraint due to the domain of the logarithmic utility function is
\begin{equation}\label{constlcbdc}
    R^d d^l + R^m m^l + {s}_{\mathrm{min}} > 0 \,.
\end{equation}

Appendix \ref{sec:app:mod:focs} reports the derived optimality conditions for HFL and LFL agents in the two economies, equations (\ref{foca})-(\ref{focml}). Since the equilibrium solutions cannot be derived analytically, I solve the model through numerical optimization where the gradient and the Hessian are provided by automatic differentiation. However, certain limits can still be explored analytically.

For instance, the optimal allocation from a pure liquidity perspective can be derived by combining the first-order conditions for deposits and CBDC in the CBDC economy. I use the conditions for the LFL agent. Using the corresponding conditions from the HFL agent's problem and following the same steps yields identical results.

Taking the difference between the first-order conditions for CBDC and deposits gives:\footnote{
See Appendix \ref{sec:app:mod:focs} for details.
}
\begin{align}
     0 = \beta (R^d - R^m) \mathcal{B} + \gamma \bigg [ \frac{(d^l)^{-\sigma}}{(d^l)^{1-\sigma} + \lambda( m^l)^{1-\sigma}} - \lambda \frac{(m^l)^{-\sigma}}{( d^l)^{1-\sigma} + \lambda(m^l)^{1-\sigma}} \bigg ],
\end{align}
where
$$\mathcal{B} = \frac{p^\epsilon}{R^d d^l + R^m m^l +{s}_{\mathrm{max}}} + \frac{1-p^\epsilon}{R^d d^l + {s}_{\mathrm{min}}}.$$
Rearranging gives
\begin{align}
     -\frac{\beta}{\gamma} (R^d - R^m) \mathcal{B} = \frac{(d^l)^{-\sigma} - \lambda(m^l)^{-\sigma}}{(d^l)^{1-\sigma} + \lambda(m^l)^{1-\sigma}}. \label{optliq}
\end{align}
By taking the limit of the liquidity preference parameter $\gamma \to \infty$, the left-hand side of equation (\ref{optliq}) vanishes. This implies that the numerator on the right-hand side must also vanish:
\begin{align}
     (d^l)^{-\sigma} = \lambda(m^l)^{-\sigma}.
\end{align}
Rearranging, the following condition is derived:
\begin{align}
     \frac{m^l}{d^l} = \lambda^{\frac{1}{\sigma}} \label{optl}.
\end{align}
This ratio between CBDC and deposits represents the allocation agents approach when considering just their liquidity needs (i.e., in the $\gamma \to \infty$ limit), and represents the upper limit of CBDC holdings.

\subsection{Calibration}\label{seccal}
I calibrate the model over an agent's total working period of $40$ years. The first half of the two-period economy is the first period in which the agent receives the endowment. The second period is the remaining half of the working life, in which the agent receives the income. Table \ref{cal_params} summarizes the baseline calibration.

The endowment $y$ is set to $1$. The discount factor is set to $\beta = 0.82$, which implies an annual discount factor of approximately $0.99$.

The risk-free return on deposits is set at the standard value of $R^d=1/\beta$, approximately equal to $1.22$, and matching an annual return of $1.01$. As mentioned previously, I assume that CBDC pays a lower return than deposits, and set $R^m = 1.10$, implying an annual return of $1.0048$. The calibration of risky asset returns and probabilities uses the binomial tree security pricing model [see, e.g., \textcite{Petters2016}]. The high return on risky assets is set to $3.77$ with a probability of $0.92$, and the low return is $0.83$ with complementary probability. See Appendix \ref{sec:app:cal} for details.

The liquidity service parameter $\gamma$ is calibrated internally to $0.05$ to match the empirical average deposit-to-consumption ratio in the SHIW data. The benefit of CBDC is normalized to $\lambda=1$ such that agents perceive CBDC and deposits as equally useful. Finally, I set the inverse elasticity of substitution between CBDC and deposits, $\sigma$, to $1/3$, which corresponds to a low degree of substitutability following \textcite{Bacchetta2025}.

\begin{table}[!htbp]
\centering
\caption{Model parameters}
\begin{threeparttable}
\begin{tabular}{l*{4}{c}}
\toprule \toprule
\multicolumn{1}{c}{Parameter} & \multicolumn{1}{c}{Value} & \multicolumn{1}{c}{Source/Motivation} & \multicolumn{1}{c}{Description} \\
\midrule 
$y$          & $1$       & Assumption & Initial endowment \\
$\beta$      & $0.82$    & Compound. Annual value $0.99$ & Discount factor \\
\midrule
$R^d$        & $1.22$     & Compound. Annual return $1.01$ & Deposits return \\
$R^m$        & $1.10$     & Assumption ($R^m < R^d$)       & CBDC return \\
$\bar{R}^a$  & $3.77$    & See Appendix \ref{sec:app:cal} & High risky assets return \\
$\ubar{R}^a$ & $0.83$    & See Appendix \ref{sec:app:cal} & Low risky assets return \\
$p^a$        & $0.92$    & See Appendix \ref{sec:app:cal} & Probability of high return \\
\midrule
$\gamma$     & $0.05$    & Internally calibrated & Liquidity service parameter \\
$\lambda$    & $1$       & Assumption ($=$ to deposits) & CBDC benefit \\
$\sigma$     & $1/3$     & \textcite{Bacchetta2025} & Inv. elasticity of substitution \\
\bottomrule
\end{tabular}
\end{threeparttable}

\vskip 0.5em
\begin{minipage}{\textwidth}
    \footnotesize
    \textit{Note}: This table reports the calibration over the period length of $20$ years.
\end{minipage}
\label{cal_params}
\end{table}

\subsection{Portfolio choices}
This section examines how high- and low-financially literate agents allocate their portfolios in different economic environments. Specifically, I study portfolio reallocation in the CBDC economy. Recall that after the introduction of CBDC, the liquidity term in the agents' utility function depends on deposits and CBDC, as specified by equation (\ref{liqserv}).

From equation (\ref{optl}), using the calibrated value for equal benefits for CBDC and deposits (i.e., $\lambda=1$), the optimal allocation of CBDC relative to the total of liquid assets is
\begin{align}
    \frac{m^j}{d^j + m^j} = 50\%. \label{liqopt}
\end{align}

This value, derived from a pure liquidity perspective, represents the upper limit for CBDC allocation and provides a useful benchmark for analyzing portfolio allocation decisions under both deterministic and stochastic second-period income scenarios.

\subsubsection{Deterministic second-period income}\label{detscenario}
I start by analyzing the agents' portfolio decisions in the deterministic case when the maximum and minimum second-period incomes are equal, i.e., $s_{\mathrm{max}} = s_{\mathrm{min}} \equiv s$. Note that the HFL agent remains subject to stochastic returns on risky assets. Following the introduction of CBDC, portfolio allocation decisions adjust according to the relative attractiveness of the digital currency.

\begin{figure}[!htbp]
\centering
\caption{Portfolio allocation with deterministic income}
\begin{minipage}{\textwidth}
    \centering
    \makebox[\textwidth]{\includegraphics[width=0.8\textwidth]{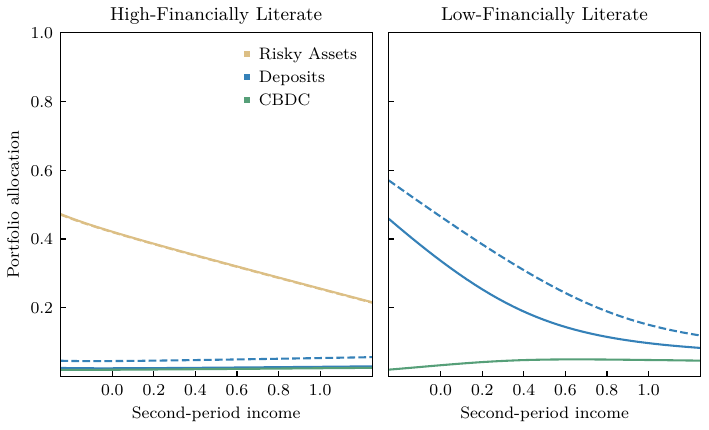}}
\end{minipage}
\vskip 0.5em
\begin{minipage}{\textwidth}
    \footnotesize
    \textit{Note}: This figure shows the portfolio allocation for high- and low-financially literate agents when the maximum and minimum second-period incomes are identical. Dashed lines represent pre-CBDC allocations, while solid lines show allocations after CBDC introduction. Values are relative to the first-period endowment $y=1$.
\end{minipage}
\label{fig:portdet}
\end{figure}

Figure \ref{fig:portdet} illustrates how agents allocate their initial endowment, $y=1$, in their portfolios as a function of second-period income values. Dashed lines represent pre-CBDC allocations, while solid lines show allocations after CBDC introduction.

In scenarios where second-period income is significantly low or even negative, both agents respond by increasing their savings, achieved by reducing current consumption to ensure sufficient funds for second-period consumption. The HFL agent invests in risky assets, as their expected return is greater than the deposit return, while the LFL agent can only save in deposits. After the CBDC introduction, the LFL agent gains access to an additional savings device but continues to allocate mainly to deposits due to their higher return relative to CBDC.

For the HFL agent, the higher return on risky assets results in higher wealth in the second period, allowing them to meet both consumption and liquidity needs. After the CBDC introduction, the HFL agent allocates around $45\%$ of liquid assets in CBDC. This value is relatively stable across all income levels and close to the optimum from a pure liquidity perspective, equation (\ref{liqopt}). In contrast, the LFL agent, who cannot invest in risky assets, must rely on the higher-return saving option (i.e., deposits) to sustain their second-period consumption. Consequently, the LFL agent prioritizes deposits over CBDC, resulting in a portfolio with a CBDC allocation that does not exceed $36\%$ across income levels, further diverging from the optimal liquidity allocation compared to the HFL agent. Note that for most income values the LFL agent's reduction in deposits is greater than the shift to CBDC, so they substitute from deposits to CBDC less than one-for-one. This is because holding CBDC is costly due to its lower return.

\begin{figure}[!htbp]
\centering
\caption{Portfolio composition with deterministic income after CBDC introduction}
\begin{minipage}{\textwidth}
    \centering
    \makebox[\textwidth]{\includegraphics[width=0.8\textwidth]{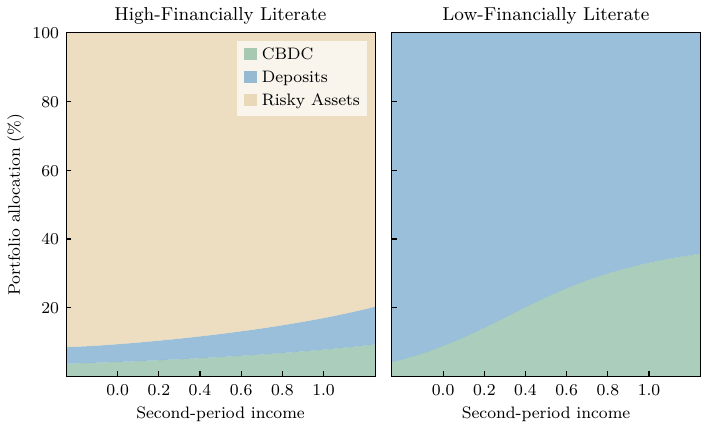}}
\end{minipage}
\vskip 0.5em
\begin{minipage}{\textwidth}
    \footnotesize
    \textit{Note}: This figure shows the portfolio composition for high- and low-financially literate agents with deterministic income after CBDC introduction.
\end{minipage}
\label{fig:portdetm}
\end{figure}

Despite the LFL agent’s preference for deposits across different second-period income values, they allocate more wealth to CBDC than the HFL agent both in absolute terms, as shown in Figure \ref{fig:portdet}, and in relative terms within their total portfolio value, with the HFL agent allocating from $4\%$ to $9\%$ to CBDC and the LFL agent ranging CBDC allocation from $4\%$ to $36\%$ depending on second-period income values, as illustrated in Figure \ref{fig:portdetm}.

This pattern qualitatively aligns with the empirical findings in Section \ref{empresult}, where low-financially literate households show a greater probability of participating following the introduction of a new financial instrument than high-financially literate households.\footnote{
The empirical analysis examines the extensive margin of participation following the introduction of the new financial instrument, while the theoretical model investigates agents' portfolio allocation decisions once the new instrument has been introduced.
} 
In the model, this difference arises because the HFL agent uses risky assets to safeguard against income risk. Consequently, this distinct portfolio behavior illustrates how financial literacy shapes not only access to assets but also the extent of participation in CBDC.

\subsubsection{Stochastic second-period income}
Next, I relax the assumption on the outcomes of the stochastic second-period income, allowing for the maximum and minimum income values to differ, i.e., $s_{\mathrm{max}}\neq s_{\mathrm{min}}$. Variation in second-period income introduces a common layer of uncertainty that affects agents' future resources, allowing me to investigate how agents with different levels of financial literacy adjust their portfolio choices, including CBDC demand, when facing such uncertainty. Specifically, I consider a maximum positive second-period income of $s_{\mathrm{max}} = 1.25$ and explore variation in the minimum income $s_{\mathrm{min}}$. As a reference, setting  $s_{\mathrm{min}} = -0.25$ implies a coefficient of variation in the income distribution of approximately $1.5$, matching the empirical coefficient of variation observed in SHIW earnings data.\footnote{
The coefficient of variation measures relative variability within a distribution, calculated as the ratio of the standard deviation to the mean.
}

\begin{figure}[!htbp]
\centering
\caption{Portfolio allocation with stochastic income}
\begin{minipage}{\textwidth}
    \centering
    \makebox[\textwidth]{\includegraphics[width=0.8\textwidth]{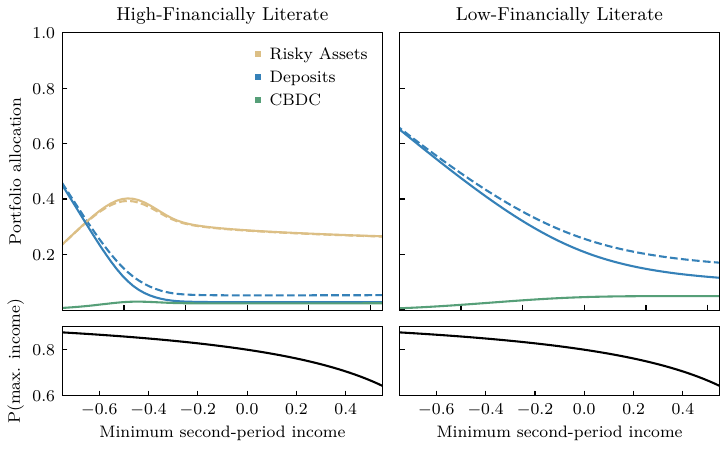}}
\end{minipage}
\vskip 0.5em
\begin{minipage}{\textwidth}
    \footnotesize
    \textit{Note}: This figure shows the portfolio allocation for high- and low-financially literate agents with stochastic income. The dashed lines represent pre-CBDC allocations, while solid lines show allocations after CBDC introduction. Variation results from adjusting the minimum income while holding the expected income constant by adjusting the probability of positive income, as shown in the lower panel. Values are relative to the first-period endowment $y=1$.
\end{minipage}
\label{fig:portstoc}
\end{figure}

Figure \ref{fig:portstoc} illustrates how agents allocate their initial endowment, $y=1$, in their portfolios as a function of minimum second-period income values. Dashed lines represent the pre-CBDC allocations, while solid lines show allocations following the CBDC introduction.

To explore the effects of income uncertainty, I vary the minimum income $s_{\mathrm{min}}$ while keeping the expected second-period income constant at $\mathbb{E}[\epsilon] = 1$ by adjusting the probability of positive income $p^\epsilon$ as follows:
\begin{align}
    p^\epsilon(s_\mathrm{min}) = \frac{\mathbb{E}[\epsilon] - s_\mathrm{min}}{s_\mathrm{max} - s_\mathrm{min}}. \label{peps}
\end{align}

This approach allows me to analyze the effect of uncertainty alone, as the expected income remains fixed and equal to $1$. This is in contrast with the deterministic scenario in Figure \ref{fig:portdet}, where I vary the expected income (i.e., $\mathbb{E}[\epsilon] = s$). As the minimum income becomes more negative (i.e., moving left in Figure \ref{fig:portstoc}), income variance increases, where the variance is
\begin{align}
    \mathrm{Var}(\epsilon) = p^\epsilon s_\mathrm{max}^2 + (1-p^\epsilon) s_\mathrm{min}^2 - \mathbb{E}^2[\epsilon].
\end{align}
Substituting equation (\ref{peps}), income variance as a function of the minimum income becomes
\begin{align}
    \mathrm{Var}(\epsilon) = (s_\mathrm{max} - \mathbb{E}[\epsilon])^2 \frac{\mathbb{E}[\epsilon] - s_\mathrm{min}}{s_\mathrm{max} - s_\mathrm{min}} + (s_\mathrm{min} - \mathbb{E}[\epsilon])^2 \frac{\mathbb{E}[\epsilon] - s_\mathrm{min}}{s_\mathrm{max} - s_\mathrm{min}}.
\end{align}
Thus, the variance increases with more negative values of the minimum income.

This income uncertainty leads to deviations from the optimal allocation considered purely from a liquidity perspective, as in the deterministic scenario. Note that the key determinant for agents' portfolio adjustments is not the mean or variance of income but rather the severity of the minimum income (i.e., the worst-case scenario agents may face). This is because agents must uphold a positive level of consumption in the second period, as captured in the constraints shown in equations (\ref{consth}), (\ref{constl}), (\ref{consthcbdc}), and (\ref{constlcbdc}). Agents ultimately reach a stable configuration for liquid asset allocation beyond a particular threshold for the minimum income, with the HFL agent achieving stability at a less negative income level.

For the HFL agent, the risky asset allocation is a hump-shaped function of the minimum income. For moderate values of $s_\mathrm{min}$, risky assets provide a hedge against low-income scenarios, but as the minimum income becomes more negative, a tipping point occurs where risk becomes too high, causing the HFL agent to shift toward safer assets.

\begin{figure}[!htbp]
\centering
\caption{Portfolio composition with stochastic income after CBDC introduction}
\begin{minipage}{\textwidth}
    \centering
    \makebox[\textwidth]{\includegraphics[width=0.8\textwidth]{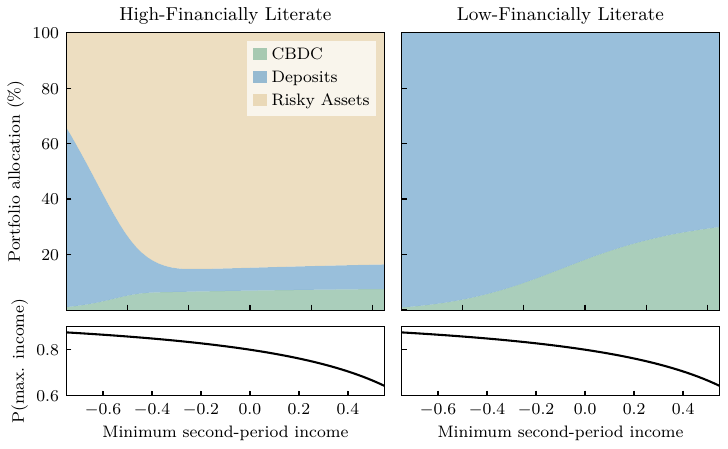}}
\end{minipage}
\vskip 0.5em
\begin{minipage}{\textwidth}
    \footnotesize
    \textit{Note}: This figure shows the portfolio composition for high- and low-financially literate agents with stochastic income after CBDC introduction.
\end{minipage}
\label{fig:portstocm}
\end{figure}

In summary, as income uncertainty increases, the financial behavior of the agents converges: the HFL agent reallocates away from risky assets toward safer, liquid instruments. Nevertheless, even with stochastic income, the LFL agent holds more CBDC than the HFL agent in absolute and relative terms, as illustrated in Figure \ref{fig:portstocm}.

\subsection{Sensitivity analysis of CBDC parameters}
Lastly, I explore the sensitivity of the model results to changes in key parameters related to CBDC.

Starting with the inverse elasticity of substitution between CBDC and deposits, $\sigma$, changes in this parameter are straightforward to interpret. In the baseline case with a high value of $\sigma$ (i.e., $\sigma=1 / 3$), CBDC and deposits are less substitutable, and are considered complements in the household's portfolio. Consequently, the household prefers to hold both assets to benefit from their combined utility rather than relying on just one. In contrast, for low values of $\sigma$ [e.g., $\sigma=1 / 20$ as considered by \textcite{Bacchetta2025}], CBDC and deposits are highly substitutable, and the household holds more of the asset with the higher return.

Next, consider the parameter capturing the relative preference for CBDC, $\lambda$. In the baseline, I assume that agents perceive CBDC and deposits as equally useful (i.e., $\lambda = 1$). A higher $\lambda$ increases the perceived benefit of CBDC, and it could be interpreted, for example, as a flight to safety toward CBDC during stress periods, reflecting its safer nature as a central bank liability, or as an increase in the liquidity benefits that CBDC provides as a means of payment. As a result, the marginal utility of holding CBDC increases, and agents shift their portfolios toward CBDC, all else equal. Conversely, a lower $\lambda$ reduces the relative attractiveness of CBDC and shifts portfolios toward deposits.

\begin{figure}[!htbp]
\centering
\caption{Portfolio allocation for the low-financially literate agent with deterministic income as a function of CBDC return}
\begin{minipage}{\textwidth}
    \centering
    \makebox[\textwidth]{\includegraphics[width=0.8\textwidth]{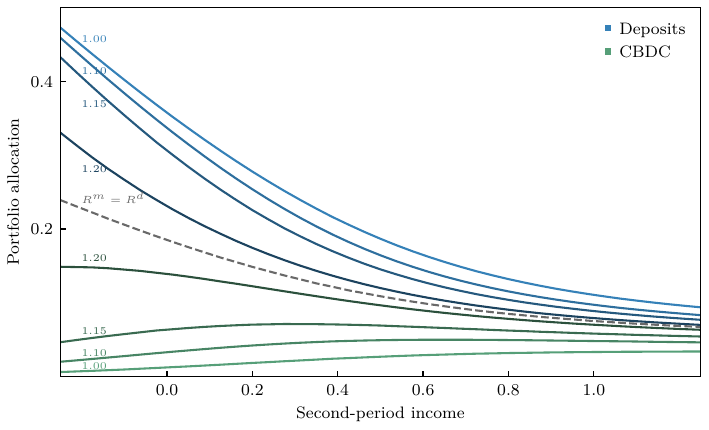}}
\end{minipage}
\vskip 0.5em
\begin{minipage}{\textwidth}
    \footnotesize
    \textit{Note}: This figure shows the portfolio allocation in the CBDC economy for the low-financially literate agent when the maximum and minimum second-period incomes are identical. Model solutions are presented for different values of CBDC return, $R^m$, equal to $1.00, 1.10, 1.15$, and $1.20$, with light color lines corresponding to lower returns. The dashed line indicates the case for $R^m = R^d \simeq 1.22$, where the allocations for CBDC and deposits coincide. Values are relative to the first-period endowment $y=1$.
\end{minipage}
\label{fig:portdetrm}
\end{figure}

The baseline model assumes a positive yet return-dominated CBDC rate of $R^m =1.10$. I vary the CBDC return and show how agents' optimal liquidity allocations shift as a result. Figure \ref{fig:portdetrm} shows the changes in the LFL agent's portfolio allocations after CBDC introduction as a function of different CBDC return values. The figure presents model solutions for CBDC returns ranging from the limit case of non-interest-bearing CBDC (i.e., $R^m = 1.00$), represented by the lightest color, to the case where the CBDC return approaches the deposit return (i.e., $R^m = 1.20$), shown with the darkest color. Even with a non-interest-bearing CBDC, the LFL agent allocates a positive share of wealth to CBDC, reflecting its value as a liquid asset, and the qualitative results hold. As the CBDC rate gets closer to the deposit rate, the optimal allocation for CBDC and deposits converge, and in the limit where the rates are equal (i.e., $R^m=R^d \simeq 1.22$), the allocations coincide, as indicated by the dashed line.

For low values of second-period income (e.g., $s=0$), the elasticities of CBDC and deposit shares to changes in CBDC return are approximately $0.63$ and $-0.63$, respectively, within the range of CBDC returns considered. The fact that the two elasticities nearly sum to zero indicates that there are no significant changes in first-period consumption. However, as second-period income increases to unity, this is no longer the case, with the elasticities of CBDC and deposit shares becoming $0.18$ and $-0.16$, respectively, and first-period consumption decreases as the CBDC return increases. The portfolio behavior of the HFL agent, not shown, follows a similar pattern to that of the LFL agent, with their allocation to risky assets remaining invariant to changes in the CBDC return.\footnote{
In the case of stochastic second-period income, agents' optimal liquidity allocations shift in a qualitatively similar way as the CBDC return changes, as illustrated in Figure \ref{fig:portstocrm} in Appendix \ref{sec:app:addfig}.
}

\section{Conclusion} \label{conclusion}
Government policies have the potential to increase participation in financial markets, which is judged to be inefficiently low. This paper evaluates the effects of government policies targeted at households on market participation, considering the role of financial literacy, with a focus on the potential introduction of a retail CBDC. By using the introduction of retail Treasury bonds in Italy as a proxy for CBDC adoption, I provide insights into how financial literacy influences households' likelihood to engage with the new instrument.

The empirical findings reveal that financial literacy positively impacts household propensity to participate after the introduction of the new policy. Interestingly, the effect is non-monotonic, with households exhibiting some but low financial literacy being more likely to participate compared to the financially illiterate and highly literate groups. The results also suggest that households reallocate their portfolios differently depending on their financial literacy following the new policy, shifting from other securities and stocks to Treasury bonds, highlighting the key role of financial literacy in reshaping household financial behavior.

Building on these empirical insights, I construct a theoretical model to explore how financial literacy impacts CBDC demand through portfolio allocation decisions. The model captures two main economic motives behind CBDC adoption: a liquidity motive, whereby agents derive utility from holding liquid assets (deposits and CBDC), and a risk-sharing motive, whereby the high-financially literate agent can use risky assets to safeguard against income risk. These channels interact with households' financial literacy, which determines access to risky investments. The low-literacy agent, limited in their access to risky investment options, relies more on liquid safe assets, while the high-literacy agent adjusts their portfolio by balancing risk and liquidity needs. This allocation pattern, similar to the empirical finding, underscores how households with some but low financial literacy may be more likely to participate in the CBDC market.

In the theoretical model, CBDC is interest-bearing but return-dominated by deposits. Although central banks are currently considering non-interest-bearing CBDC designs, the sensitivity analysis shows that as long as CBDC remains return-dominated, the core results are robust, including in the case of a non-interest-bearing CBDC. The model is intentionally stylized and abstracts from design and equilibrium features that are currently discussed in the policy debate on CBDCs. Deposit returns are taken as exogenous, while in a richer environment, banks could respond to CBDC introduction by adjusting deposit rates. Additionally, to mitigate unintended effects on bank intermediation, many proposals consider quantitative limits on individual CBDC holdings [see, e.g., \textcite{ecb2025}]. Such limits are not modeled here; in practice, they would make CBDC operate mainly as a transactional instrument rather than a store of value. Privacy and other non-pecuniary attributes are also not modeled explicitly and are instead captured in reduced form by households’ relative preference for CBDC liquidity.

This paper offers valuable insights for policymakers. First, it highlights the importance of considering financial literacy when designing a CBDC. Since households with different levels of financial literacy respond differently to new financial instruments, CBDC communication should be clear and inclusive so that information and onboarding are feasible across the full distribution of financial literacy. Second, the model provides insights into how different agents might reallocate their portfolios following the introduction of CBDC, depending on design features and the overall economic environment.

Future research could extend this work by using actual CBDC data when available, studying adoption dynamics over longer horizons, and further exploring the behavioral aspects of household financial decision-making in the context of new financial instruments. Additionally, an interesting extension would be to incorporate payment-specific adoption forces, such as network effects, which can generate strategic complementarities in technology adoption as highlighted in recent work on P2P digital payments [see, e.g., \textcite{Alvarez2023}]. These extensions would offer a richer understanding of the dynamics at play and provide more targeted recommendations for policy design.

\newpage

\nocite{*}
\printbibliography[
    heading=bibintoc,
    title=References,
    ]
\newpage
\appendix

\section{Additional material} \label{sec:app:additional}
\renewcommand{\thetable}{A.\arabic{table}}
\renewcommand{\thefigure}{A.\arabic{figure}}
\setcounter{table}{0}
\setcounter{figure}{0}
\begin{table}[!htbp]
\centering
\caption{Italian government securities}
\begin{threeparttable}
\begin{tabular}{l p{4cm} p{2cm} p{2cm} p{2cm}}
\toprule\toprule
& \textit{BOTs} & \textit{CTZs} & \textit{CCTs} & \textit{BTPs}\\
\midrule
Maturity & $<$ 1 year & 2 years & 5-7 years & 3-50 years \\
Coupon & \multicolumn{2}{c}{Discount at issuance} & \multicolumn{2}{c}{Floating semi-annual} \\
Auction type & Competitive yield & \multicolumn{3}{c}{Discretional choice of price/quantity issued} \\
\bottomrule
\end{tabular}
\end{threeparttable}
\vskip 0.5em
\begin{minipage}{\textwidth}
    \footnotesize
    \textit{Note}: This table compares the different Italian government securities. \textit{Buoni del Tesoro Ordinari (BOTs)} correspond to Treasury bills, \textit{Certificato di Credito del Tesoro (CCTs)} correspond to Treasury certificates, \textit{Certificato del Tesoro Zero-coupon (CTZs)} correspond to Zero-coupon bonds, and \textit{Buoni del Tesoro Poliennali (BTPs)} correspond to Treasury bonds with long-term maturity. All four securities have a minimum denomination of €$1,000$ and are subject to a $12.5\%$ tax rate.  \\
    \textit{Source}: Italian Ministry of Economy and Finance.
\end{minipage}
\label{bondcat}
\end{table}

\subsection{Subsequent retail Treasury bond issuances} \label{sec:app:additional:valore}
The first issue of retail Treasury bond \textit{Valore} in June 2023 attracted over $18.1$ billion euros with approximately $650,000$ contracts; the second issue in October 2023 raised $17.2$ billion euros with around $642,000$ contracts; and the third issuance in February 2024 ended with $18.3$ billion euros and $656,369$ contracts, marking the highest result ever recorded in terms of the amount subscribed and contracts concluded in a single issuance of government bonds dedicated to retail investors [see \textcite{mef2023a}, \textcite{mef2023b}, and \textcite{mef2024}]. The fourth issuance in May 2024 saw a decline in the public response, with $11.2$ billion euros and $384,295$ contracts. This lesser response may be attributed to the anticipated cut in key interest rates by the European Central Bank.

In February 2025, the Italian government introduced a new retail Treasury bond within the \text{Valore} family, called \textit{BTP Più}. This bond has quarterly coupon payments and maintains the ``step-up'' mechanism but has a longer maturity of 8 years. Its innovation is the option for early redemption after 4 years: Investors who purchase the bond during the placement period and hold it until the exercise date in 2029 can redeem their principal, fully or partially, in increments of at least €$1,000$. Those who do not exercise this option will continue to receive step-up coupons and are guaranteed the full principal at maturity. The first issue of \textit{BTP Più} raised approximately $15$ billion euros, a couple of billion less than the record of Treasury bonds \textit{Valore}, but four above the previous issue, with around $642,000$ contracts [\textcite{mef2025}].

\subsection{Sampling weights} \label{sec:app:additional:weights}
In the 2020 SHIW survey, a new sampling scheme was introduced to improve the quality of estimators for economic analysis. This structural change requires the adoption of specific weighting techniques for historical comparison with previous editions. More specifically, sampling weights were constructed using a statistical rebalancing technique [i.e., ranking, see \textcite{Kalton2003}] that assigns each household interviewed in 2020 the same probability of being interviewed as under the previous design. Throughout the paper, all the descriptive statistics are computed using the sampling weights.

\subsection{SHIW financial literacy questions} \label{sec:app:additional:shiwq}
Financial literacy questions are asked in the 2006, 2008, 2010, 2016, and 2020 SHIW waves. The correct answer is highlighted among those proposed.

\subsubsection{Interests}
\textit{Suppose you put $100$ euros into a (no fee, tax-free) savings account with a guaranteed interest rate of $2\%$ per year. You don’t make any further payments into this account and you don’t withdraw any money. How much would be in the account at the end of $5$ years, once the interest payment is made?}\footnote{In the 2006 wave, the question is slightly different, considering an amount equal to $1,000$ euros and 2 years. In this case, the answer ``No answer'' is not available.}
\textit{\begin{enumerate}
    \item Less than $102$ euro
    \item Exactly $102$ euro
    \item \textbf{More than 102 euro}
    \item Don't know
    \item No answer
\end{enumerate}}

\subsubsection{Inflation}
\textit{Suppose you put $1,000$ euros into a (no fee, tax free) savings account with a guaranteed interest rate of $1\%$ per year. Suppose furthermore inflation stays at $2$\%. In one year’s time will you be able to buy the same amount of goods that you could buy by spending today $1,000$ euros?}
\textit{\begin{enumerate}
    \item Yes
    \item \textbf{No, less than I could buy today}
    \item No, more than I could buy today
    \item Don't know
    \item No answer
\end{enumerate}}

\subsubsection{Risk diversification}
\textit{In your opinion, the purchase of shares of one company usually provides a safer return than buying shares of a wide range of companies through a mutual fund?}\footnote{In the 2008 and 2010 waves, the question is formulated as: \textit{Which of the following investment strategies do you think entails the greatest risk of losing your capital? (1) \textbf{Investing in the shares of a single company}, (2) Investing in the shares of more than one company, (3) Don't know, (4) No answer}. In the 2008 wave, the answer ``No answer'' is not available.}
\textit{\begin{enumerate}
    \item True
    \item \textbf{False}
    \item Don't know
    \item No answer
\end{enumerate}}

\subsubsection{Mortgages}
\textit{Which of the following types of mortgage do you think would allow you from the very start to fix the maximum amount and number of installments to be paid before the debt is extinguished?}\footnote{In the 2006 and 2008 waves, the answer ``No answer'' is not available.}
\textit{\begin{enumerate}
    \item Floating-rate mortgage
    \item \textbf{Fixed-rate mortgage}
    \item Floating-rate mortgage with fixed installments \item Don't know
    \item No answer
\end{enumerate}}

\begin{table}[!htbp]
\centering
\caption{Correct responses to financial literacy questions}
\begin{threeparttable}
\begin{tabular}{lcccc}
\toprule\toprule
\multirow{2}{*}{} & \multicolumn{1}{c}{2008} & \multicolumn{1}{c}{2010} & \multicolumn{1}{c}{2016} & \multicolumn{1}{c}{2020} \\
\midrule
Inflation & 72.84 & 69.55 & 61.42 & 59.29 \\
Risk diversification & 26.47 & 20.79 & 51.62 & 54.51 \\
Interest & - & - & 50.12 & 49.59 \\
Mortgages & 65.97 & 58.38 & - & - \\
\midrule
Average & 55.09 & 49.57 & 54.39 & 54.47 \\
\bottomrule
\end{tabular}
\end{threeparttable}

\vskip 0.5em
\begin{minipage}{\textwidth}
    \footnotesize
    \textit{Note}: This table reports the aggregate percentage of correct responses to financial literacy questions. Note that incorrect responses include ``Don't know'' and ``No answer''. The average refers to the average number of correct answers. \\
    \textit{Source}: SHIW and author calculations.
\end{minipage}
\label{tab:correct}
\end{table}

\subsection{Descriptive statistics} \label{sec:app:additional:sumstat}

\begin{figure}[!htbp]
\centering
\caption{Financial literacy score across Italian regions}
\begin{minipage}{\textwidth}
    \centering
    \makebox[\textwidth]{\includegraphics[width=0.4\textwidth]{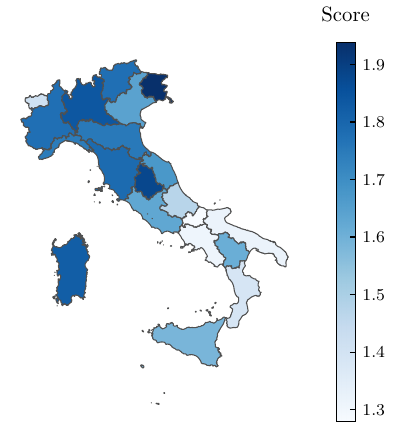}}
\end{minipage}
\vskip 0.5em
\begin{minipage}{\textwidth}
    \footnotesize
    \textit{Note}: This map shows the average financial literacy score across Italian regions. The full range of possible scores goes from 0 to 3. \\
    \textit{Source}: SHIW and author calculations.
\end{minipage}
\label{fig:geomap}
\end{figure}

Table \ref{stats} shows the statistics for the financial literacy score across various demographic groups, highlighting disparities based on factors such as education, age, gender, marital status, employment, and risk aversion. The mean score increases significantly with higher educational qualifications, while the increase in the population of the municipality of residence is relatively modest. Financial literacy scores tend to rise with age but decrease for individuals over 65, consistent with existing literature [e.g., \textcite{Lusardi2022}]. Males, on average, have higher scores than females, and married individuals score higher than single individuals. Employment status also appears to matter, with employed individuals showing higher scores on average than those not employed. Additionally, the mean score is decreasing in risk aversion. These statistics highlight the disparities in financial literacy across different demographic groups. Figure \ref{fig:geomap} further illustrates these disparities by showing the regional distribution of financial literacy scores in Italy. Regions in the North generally exhibit higher scores than those in the South, with overall variation in scores ranging from $1.3$ to $1.9$.

\begin{table}[!htbp]
\centering
\caption{Financial literacy score by demographics}
\begin{threeparttable}
\begin{tabular}{l*{3}{c}}
\toprule\toprule
Variable & Mean & Std & N \\
\midrule
\multicolumn{3}{l}{\textbf{Educational qualification}} & \\
None                        & 0.74 & 0.92 & 1,115 \\
Elementary school           & 1.21 & 1.02 & 6,469 \\
Middle school               & 1.65 & 0.97 & 10,025 \\
High school                 & 1.93 & 0.88 & 7,695 \\
Bachelor's degree           & 2.09 & 0.88 & 3,865 \\
Post-graduate qualification & 2.17 & 0.88 & 418 \\
\midrule
\multicolumn{3}{l}{\textbf{Age}} & \\
Up to 30     & 1.65 & 0.96 & 828 \\
31 - 40      & 1.75 & 0.92 & 2,683 \\
41 - 50      & 1.79 & 0.92 & 5,180 \\
51 - 65      & 1.82 & 0.95 & 9,262 \\
More than 65 & 1.44 & 1.07 & 11,634 \\
\midrule
\multicolumn{3}{l}{\textbf{Gender}} & \\
Male   & 1.79 & 0.96 & 17,173 \\
Female & 1.47 & 1.04 & 12,414 \\
\midrule
\multicolumn{3}{l}{\textbf{Marital status}} & \\
Married  & 1.77 & 0.96 & 17,883 \\
Single   & 1.49 & 1.05 & 11,704 \\
\midrule
\multicolumn{3}{l}{\textbf{Work status}} & \\
Employed     & 1.88 & 0.91 & 12,636 \\
Not employed & 1.49 & 1.04 & 16.951 \\
\midrule
\multicolumn{3}{l}{\textbf{Municipality Population}} & \\
Up to 5,000       & 1.59 & 1.00 & 3,317 \\
5,000 - 20,000    & 1.60 & 0.97 & 3,962 \\
20,000 - 50,000   & 1.66 & 1.00 & 8,584 \\
50,000 - 200,000  & 1.68 & 1.02 & 9,222 \\
More than 200,000 & 1.70 & 1.00 & 4,502 \\
\midrule
\multicolumn{3}{l}{\textbf{Risk aversion}} & \\
Low    & 1.83 & 0.98 & 311 \\
Medium & 1.78 & 0.96 & 13,601\\
High   & 1.55 & 1.03 & 15,675 \\
\bottomrule
\end{tabular}
\end{threeparttable}
\vskip 0.5em
\begin{minipage}{\textwidth}
    \footnotesize
    \textit{Note}: This table reports the mean and standard deviation of the financial literacy score by demographic variables, where the range of possible scores goes from 0 to 3. Single includes singles, divorced, and widowed individuals. Employed respondents include employees and self-employed individuals. \\
    \textit{Source}: SHIW and author calculations.
\end{minipage}
\label{stats}
\end{table}

\begin{figure}[!htbp]
\centering
\caption{Wealth distribution across financial literacy scores}
\begin{minipage}{\textwidth}
    \centering
    \makebox[\textwidth]{\includegraphics[width=0.8\textwidth]{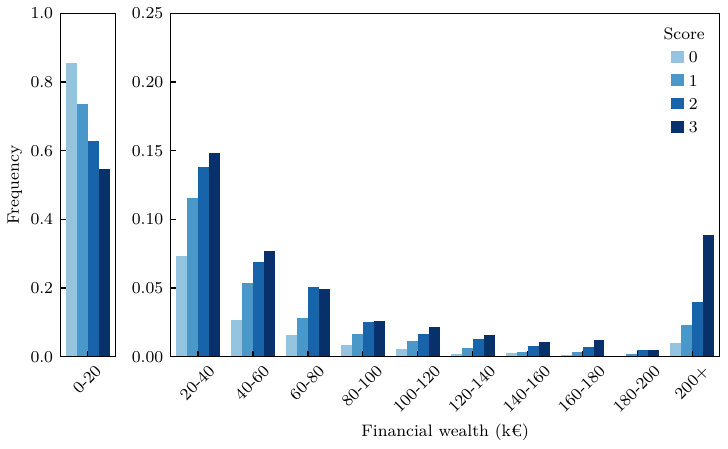}}
\end{minipage}
\vskip 0.5em
\begin{minipage}{\textwidth}
    \footnotesize
    \textit{Note}: This figure shows the wealth distribution for each financial literacy score level. Wealth is measured as total financial wealth, categorized into intervals of €20,000. The distribution reflects the average over the sample. \\
    \textit{Source}: SHIW and author calculations.
\end{minipage}
\label{fig:finlitwealth}
\end{figure}

In addition to demographic and regional differences, another important dimension of financial literacy is its relationship with financial wealth. Figure \ref{fig:finlitwealth} shows the wealth distribution across financial literacy scores. Households in the upper wealth brackets present higher financial literacy scores (2 and 3), while households in the lower wealth groups, particularly in the 0–20 k€ range illustrated in the left panel, show lower financial literacy scores (0 and 1). This suggests a positive correlation between financial literacy and wealth accumulation.

\section{Empirical results} \label{sec:app:emp}
\renewcommand{\thetable}{B.\arabic{table}}
\renewcommand{\thefigure}{B.\arabic{figure}}
\setcounter{table}{0}
\setcounter{figure}{0}
\subsection{Logit odds ratios} \label{sec:app:emp:odds}
To provide an intuition for interpreting the estimated coefficient of interest $\hat{\beta}$, equation (\ref{baseline}) can be rewritten as:
\begin{align}
    z = \hat{\beta} ({D}^{12} \times {LS}^{10}) + \mathcal{A}, \quad \mathcal{A} = \hat{\alpha} + \sum_{\ell=1}^{3} \hat{\gamma}_\ell {LS}^{10}_{\ell} + \hat{\boldsymbol{\eta}}^\intercal {\boldsymbol{X}}, \tag{B1}
\end{align}
where $\mathcal{A}$ includes the effects of the other predictors.

Recalling that the propensity to participate is estimated with a logistic function, as in equation (\ref{linkf}), the differential policy-financial literacy effects when the interaction term $D^{12} \times {LS}^{10}$ is equal to $1$ and $0$ are, respectively:
\begin{align}
    \ln \left( \frac{P_1}{1-P_1} \right) = \hat{\beta} +\mathcal{A}, \qquad \ln \left( \frac{P_0}{1-P_0} \right) = \mathcal{A}. \tag{B2}
\end{align}
By taking the difference, I obtain an expression that relates the probabilities of participation only to the coefficient of the policy-literacy term and is independent of $\mathcal{A}$:
\begin{align}
    \ln \left( \frac{P_1}{1-P_1} \right) - \ln \left( \frac{P_0}{1-P_0} \right) = \hat{\beta}. \tag{B3}
\end{align}
Taking exponents on both sides gives the odds ratio:
\begin{align}
    \frac{P_1}{1-P_1} = \frac{P_0}{1-P_0} \times \exp(\hat{\beta}). \tag{B4}
\end{align}

\begin{landscape}
\begin{table}[!htbp]
\centering
\caption{Logit estimation results}
\begin{threeparttable}
\begin{tabular}{l@{\extracolsep{-30pt}}S@{\extracolsep{-30pt}}S@{\extracolsep{-30pt}}S@{\extracolsep{-30pt}}S@{\extracolsep{-30pt}}S@{\extracolsep{-30pt}}S}
\toprule\toprule
\multirow{2}{*}{\begin{tabular}[c]{@{}l@{}}Dep. Var.: \\ Market Participation\end{tabular}} 
& \multicolumn{2}{c}{Treasury bond} 
& \multicolumn{2}{c}{Other Govt. Bond} 
& \multicolumn{2}{c}{Stocks} \\
\cmidrule(lr){2-3} \cmidrule(lr){4-5} \cmidrule(lr){6-7}
& \multicolumn{1}{c}{(1)} & \multicolumn{1}{c}{(2)} 
& \multicolumn{1}{c}{(3)} & \multicolumn{1}{c}{(4)} 
& \multicolumn{1}{c}{(5)} & \multicolumn{1}{c}{(6)} \\
\midrule
$D^{12} \times {LS}^{10}$ & 0.14\textsuperscript{***} (0.05) & & -0.21\textsuperscript{***} (0.03) & & -0.05\textsuperscript{**} (0.03) & \\
$D^{12} \times {LS}_1^{10}$ & & 0.59\textsuperscript{**} (0.26) & & -0.28\textsuperscript{**} (0.13) & & -0.24\textsuperscript{***} (0.09) \\
$D^{12} \times {LS}_2^{10}$ & & 0.22\textsuperscript{*} (0.13) & & -0.42\textsuperscript{***} (0.09) & & -0.13\textsuperscript{**} (0.06) \\
$D^{12} \times {LS}_3^{10}$ & & 0.45\textsuperscript{*} (0.26) & & -0.63\textsuperscript{***} (0.18) & & 0.05\text{ } (0.17) \\
\midrule
${LS}_1^{10}$ & 1.27\textsuperscript{***} (0.48) & 1.00\textsuperscript{*} (0.51) & 0.91\textsuperscript{***} (0.20) & 0.94\textsuperscript{***} (0.21) & 1.35\textsuperscript{***} (0.23) & 1.44\textsuperscript{***} (0.23)\\
${LS}_2^{10}$ & 1.84\textsuperscript{***} (0.46) & 1.88\textsuperscript{***} (0.46) & 1.22\textsuperscript{***} (0.20) & 1.22\textsuperscript{***} (0.20) & 1.72\textsuperscript{***} (0.22) & 1.73\textsuperscript{***} (0.22)\\
${LS}_3^{10}$ & 1.31\textsuperscript{**} (0.51) & 1.29\textsuperscript{**} (0.53) & 1.32\textsuperscript{***} (0.22) & 1.31\textsuperscript{***} (0.23) & 0.95\textsuperscript{***} (0.25) & 0.85\textsuperscript{***} (0.26)\\
Age & 0.10\textsuperscript{**} (0.04) & 0.10\textsuperscript{**} (0.04) & 0.08\textsuperscript{***} (0.03) & 0.08\textsuperscript{**} (0.03) & 0.12\textsuperscript{***} (0.02) & 0.12\textsuperscript{***} (0.02)\\
Age$^2$ & -0.00\textsuperscript{*} (0.00) & -0.00\textsuperscript{*} (0.00) & -0.00\textsuperscript{**} (0.00) & -0.00\textsuperscript{**} (0.00) & -0.00\textsuperscript{***} (0.00) & -0.00\textsuperscript{***} (0.00)\\
Married & 0.29\text{ } (0.19) & 0.29\text{ } (0.19) & 0.08\text{ } (0.11) & 0.09\text{ } (0.11) & 0.14\text{ } (0.09) & 0.14\text{ } (0.09)\\
Female  & -0.29\textsuperscript{*} (0.18) & -0.29\text{ } (0.18) & -0.02\text{ } (0.10) & -0.02\text{ } (0.10) & -0.51\textsuperscript{***} (0.09) & -0.50\textsuperscript{***} (0.09)\\
High-Risk Aversion & -1.13\textsuperscript{***} (0.18) & -1.14\textsuperscript{***} (0.18) & -0.19\textsuperscript{**} (0.09) & -0.19\textsuperscript{**} (0.09) & -1.08\textsuperscript{***} (0.09) & -1.08\textsuperscript{***} (0.09)\\
Constant & -8.10\textsuperscript{***} (1.47) & -8.05\textsuperscript{***} (1.47) & -6.45\textsuperscript{***} (0.84) & -6.44\textsuperscript{***} (0.84) & -5.79\textsuperscript{***} (0.67) & -5.81\textsuperscript{***} (0.67)\\
\midrule
Pseudo $R^2$ & 0.017 & 0.017 & 0.016 & 0.016 & 0.065 & 0.065 \\
\bottomrule
\end{tabular}
\begin{tablenotes}[flushleft]
    \footnotesize
    \item \textit{Note}: This table reports coefficient estimates using Generalized Estimating Equations (GEE), an extension of GLM for panel data. The sample counts a panel of 4,611 households over a two-year period. Robust standard errors in parentheses. Time-fixed effect included. The indicator for high risk aversion is measured in 2010.
    \item Columns (1), (3), and (5) refer to specification (\ref{baseline}), while columns (2), (4), and (6) refer to specification (\ref{level}).
    \item In specification (\ref{level}), the baseline financial literacy level is 0. * $p < 0.1$; ** $p < 0.05$; *** $p < 0.01$.
\end{tablenotes}
\end{threeparttable}

\label{logit}
\end{table}
\end{landscape}

\subsection{Post-estimation inference} \label{sec:app:emp:inference}
Let $\delta = \beta_1 - \beta_2$ represent the difference in the respective model coefficients for the low- and high-literacy households from the fitted logistic model. I set up a one-sided hypothesis
$$
H_0: \delta \leq 0 \quad \text{vs.}\quad H_1: \delta > 0.
$$
Under the null hypothesis, the effect of the low-literate is at most equal to (or possibly less than) that of the high-literate. The alternative hypothesis states that low literacy has a larger effect.

To test this, one typically uses a Wald-type $z$-statistic rooted in large-sample theory. Concretely, after estimating the model’s coefficients $\hat{\beta}_1$ and $\hat{\beta}_2$, I compute $\hat{\delta} = \hat{\beta}_1 - \hat{\beta}_2$ and obtain its estimated variance from the model's covariance matrix. Because quasi-likelihood estimators (as in the case of Generalized Estimating Equations, an extension of GLM for panel data used here) are asymptotically normal under standard regularity conditions (e.g., correct mean specification, no perfect separation, sufficiently large sample), the ratio $r = \hat{\delta} \big/ \sqrt{\text{Var}(\hat{\delta})}$ serves as the test statistic and behaves approximately as a standard normal random variable for large samples. For a one-sided test, the p-value is $p=1-\phi(r)$, where $\phi(\cdot)$ is the standard normal cumulative distribution function. A sufficiently small p-value leads to rejecting $H_0$ and concluding that the evidence supports $\beta_1 > \beta_2$. The test rejects $H_0$ at the $10\%$ significance level, indicating that the estimated coefficient of low-literacy households is statistically larger than that of high-literacy households.

\subsection{Robustness} \label{sec:app:emp:robustness}
I implement a data-driven control selection procedure based on the post-double-selection framework of \textcite{Belloni2013}. The objective is to guard against omitted variable bias when a rich set of candidate controls is available, while retaining valid inference on the coefficients of interest under approximate sparsity.

Let $Y_{it}$ denote the binary outcome, let $D_{it}$ denote the treatment variable(s) of interest (for example, the post introduction interaction term(s) used in Section \ref{empexercise}), and let $Z_{it}$ denote a high dimensional set of potential controls constructed from the SHIW covariates. The procedure consists of three steps. First, I run a penalized regression of $Y_{it}$ on $Z_{it}$ to select controls predictive of the outcome. Second, I run a penalized regression of $D_{it}$ on $Z_{it}$ to select controls predictive of treatment assignment; when $D_{it}$ contains multiple treatment terms (as in the non-linear specification equation (\ref{level})), I apply this step to each term and take the union of the selected controls. Third, I estimate the main specification by unpenalized GEE including $D_{it}$ and the union of controls selected in the first two steps, and I report robust standard errors.

In the selection steps, I use the smoothly clipped absolute deviation (SCAD) penalty of \textcite{Fan2001}, which is designed to reduce shrinkage bias for larger coefficients. Since the estimation is conducted in a correlated panel setting, the penalized selection step is implemented within the GEE framework following \textcite{Wang2011}. Final inference is based on the unpenalized GEE estimated on the union of selected controls, as in the post-double-selection approach.

The set of candidate controls mirrors the demographic and wealth variables discussed in Appendix \ref{sec:app:additional:sumstat}. Table \ref{logit_robustness} reports the logit regression estimation results using the covariates selected by the post-double-selection procedure. Compared with Table \ref{logit}, the qualitative results remain unchanged. Under the linear specification, following the introduction of retail Treasury bonds, the propensity to participate in the Treasury bond market increases with financial literacy, while that in the other government bonds and the stock market declines. Under the non-linear specification, the heterogeneity pattern across financial literacy levels is preserved.

\begin{landscape}
\begin{table}[!htbp]
\centering
\caption{Logit estimation results - Robustness}
\begin{threeparttable}
\begin{tabular}{l@{\extracolsep{-30pt}}S@{\extracolsep{-30pt}}S@{\extracolsep{-30pt}}S@{\extracolsep{-30pt}}S@{\extracolsep{-30pt}}S@{\extracolsep{-30pt}}S}
\toprule\toprule
\multirow{2}{*}{\begin{tabular}[c]{@{}l@{}}Dep. Var.: \\ Market Participation\end{tabular}}
& \multicolumn{2}{c}{Treasury bond}
& \multicolumn{2}{c}{Other Govt. Bond}
& \multicolumn{2}{c}{Stocks} \\
\cmidrule(lr){2-3} \cmidrule(lr){4-5} \cmidrule(lr){6-7}
& \multicolumn{1}{c}{(1)} & \multicolumn{1}{c}{(2)}
& \multicolumn{1}{c}{(3)} & \multicolumn{1}{c}{(4)}
& \multicolumn{1}{c}{(5)} & \multicolumn{1}{c}{(6)} \\
\midrule
$D^{12} \times {LS}^{10}$      & 0.16\textsuperscript{***} (0.06) &  & -0.23\textsuperscript{***} (0.04) &  & -0.07\textsuperscript{**} (0.03) &  \\
$D^{12} \times {LS}_1^{10}$    &  & 0.64\textsuperscript{**} (0.27)  &  & -0.29\textsuperscript{**} (0.14)  &  & -0.31\textsuperscript{***} (0.12) \\
$D^{12} \times {LS}_2^{10}$    &  & 0.26\textsuperscript{*} (0.14)   &  & -0.45\textsuperscript{***} (0.10) &  & -0.16\textsuperscript{**} (0.08) \\
$D^{12} \times {LS}_3^{10}$    &  & 0.50\textsuperscript{*} (0.28)   &  & -0.68\textsuperscript{***} (0.20) &  & 0.05\text{ } (0.20) \\
\midrule
${LS}_1^{10}$ & 0.39\text{ } (0.49) & 0.11\text{ } (0.52) & 0.34\textsuperscript{*} (0.20) & 0.37\textsuperscript{*} (0.21) & 0.55\textsuperscript{**} (0.23) & 0.67\textsuperscript{***} (0.23) \\
${LS}_2^{10}$ & 0.76\text{ } (0.47) & 0.80\textsuperscript{*} (0.47) & 0.49\textsuperscript{**} (0.20) & 0.49\textsuperscript{**} (0.20) & 0.76\textsuperscript{***} (0.22) & 0.77\textsuperscript{***} (0.22) \\
${LS}_3^{10}$ & 0.66\text{ } (0.52) & 0.65\text{ } (0.54) & 0.89\textsuperscript{***} (0.22) & 0.89\textsuperscript{***} (0.23) & 0.33\text{ } (0.25) & 0.20\text{ } (0.26) \\
Age & 0.00\text{ } (0.01) & 0.00\text{ } (0.01) & 0.01\textsuperscript{***} (0.01) & 0.01\textsuperscript{***} (0.01) & 0.00\text{ } (0.00) & 0.00\text{ } (0.00) \\
Married & 0.22\text{ } (0.19) & 0.22\text{ } (0.19) & -0.09\text{ } (0.11) & -0.09\text{ } (0.11) & 0.14\text{ } (0.10) & 0.14\text{ } (0.10) \\
Female  & -0.08\text{ } (0.19) & -0.08\text{ } (0.19) & 0.13\text{ } (0.10) & 0.13\text{ } (0.10) & -0.31\textsuperscript{***} (0.09) & -0.30\textsuperscript{***} (0.10) \\
High-Risk Aversion & -0.75\textsuperscript{***} (0.19) & -0.75\textsuperscript{***} (0.19) & 0.09\text{ } (0.10) & 0.09\text{ } (0.10) & -0.80\textsuperscript{***} (0.10) & -0.80\textsuperscript{***} (0.10) \\
Employed & -0.02\text{ } (0.21) & -0.03\text{ } (0.21) & -0.19\text{ } (0.13) & -0.19\text{ } (0.13) & 0.25\textsuperscript{**} (0.12) & 0.25\textsuperscript{**} (0.12) \\
Education & 0.17\textsuperscript{*} (0.09) & 0.17\textsuperscript{*} (0.09) & 0.05\text{ } (0.06) & 0.05\text{ } (0.06) & 0.31\textsuperscript{***} (0.05) & 0.31\textsuperscript{***} (0.05) \\
Municipality & 0.12\textsuperscript{*} (0.07) & 0.12\textsuperscript{*} (0.07) & -0.02\text{ } (0.04) & -0.02\text{ } (0.04) & -0.02\text{ } (0.04) & -0.02\text{ } (0.04) \\
Log-Wealth & 0.75\textsuperscript{***} (0.08) & 0.75\textsuperscript{***} (0.08) & 0.60\textsuperscript{***} (0.05) & 0.60\textsuperscript{***} (0.05) & 0.86\textsuperscript{***} (0.05) & 0.86\textsuperscript{***} (0.05) \\
Constant & -7.83\textsuperscript{***} (0.87) & -7.82\textsuperscript{***} (0.87) & -5.50\textsuperscript{***} (0.50) & -5.50\textsuperscript{***} (0.50) & -5.71\textsuperscript{***} (0.45) & -5.73\textsuperscript{***} (0.45) \\
\midrule
Pseudo $R^2$ & 0.05 & 0.05 & 0.08 & 0.08 & 0.21 & 0.21 \\
\bottomrule
\end{tabular}
\begin{tablenotes}[flushleft]
    \footnotesize
    \item \textit{Note}: This table reports coefficient estimates using Generalized Estimating Equations (GEE), an extension of GLM for panel data. The sample counts a panel of 4,611 households over a two-year period. Robust standard errors in parentheses. Time-fixed effect included. The two indicators for high risk aversion and employment status, as well as education level, municipality size, and log-wealth, are measured in 2010.
    \item Columns (1), (3), and (5) refer to specification (\ref{baseline}), while columns (2), (4), and (6) refer to specification (\ref{level}).
    \item In specification (\ref{level}), the baseline financial literacy level is 0. * $p < 0.1$; ** $p < 0.05$; *** $p < 0.01$.
\end{tablenotes}
\end{threeparttable}

\label{logit_robustness}
\end{table}
\end{landscape}

\section{Theoretical model} \label{sec:app:mod}
\renewcommand{\thetable}{C.\arabic{table}}
\renewcommand{\thefigure}{C.\arabic{figure}}
\setcounter{table}{0}
\setcounter{figure}{0}
\subsection{Optimal conditions} \label{sec:app:mod:focs}
In the pre-CBDC economy, the HFL agent's first-order condition for risky assets, $\partial \mathbb{E}[\mathcal{U}^h]/\partial a=0$, is:
\begin{align}
    \frac{1}{y-a-d^h} &= \beta \bigg \{ \bar{R}^a p^a \bigg [ \frac{p^\epsilon}{\bar{R}^a a + R^d d^h +{s}_{\mathrm{max}}} + \frac{1-p^\epsilon}{\bar{R}^a a + R^d d^h + {s}_{\mathrm{min}}} \bigg ] \nonumber \\
    & \hspace{1cm} + \ubar{R}^a (1-p^a) \bigg [ \frac{p^\epsilon}{\ubar{R}^a a + R^d d^h +{s}_{\mathrm{max}}} + \frac{1-p^\epsilon}{\ubar{R}^a a + R^d d^h + {s}_{\mathrm{min}}} \bigg ] \bigg \} \label{foca} \tag{C1}.
\end{align}
The first-order condition for deposits, $\partial \mathbb{E}[\mathcal{U}^h]/\partial d^h=0$, is:
\begin{align}
    \frac{1}{y-a-d^h} =& \beta R^d \bigg \{ p^a \bigg [ \frac{p^\epsilon}{\bar{R}^a a + R^d d^h +{s}_{\mathrm{max}}} + \frac{1-p^\epsilon}{\bar{R}^a a + R^d d^h + {s}_{\mathrm{min}}} \bigg ] \nonumber \\
    & \hspace{1cm} + (1-p^a) \bigg [ \frac{p^\epsilon}{\ubar{R}^a a +R^d d^h +{s}_{\mathrm{max}}} + \frac{1-p^\epsilon}{\ubar{R}^a a +R^d d^h + {s}_{\mathrm{min}}} \bigg ] \bigg \} + \frac{\gamma}{d^h}. \label{focdh} \tag{C2}
\end{align}
While the LFL agent's first-order condition, $\partial \mathbb{E}[\mathcal{U}^l]/\partial d^l=0$, is:
\begin{align}
    \frac{1}{y-d^l} =& \beta R^d \bigg [ \frac{p^\epsilon}{R^d d^l +{s}_{\mathrm{max}}} + \frac{1-p^\epsilon}{R^d d^l + {s}_{\mathrm{min}}} \bigg ] + \frac{\gamma}{d^l}. \label{focdl} \tag{C3}
\end{align}

\newpage
After the CBDC introduction, the HFL agent's first-order conditions for risky assets, deposits, and CBDC are, respectively:
\begin{align}
    a: \qquad &\frac{1}{y-a-d^h-m^h} = \beta \bigg\{ \bar{R}^a p^a \bigg [ \frac{p^\epsilon}{\bar{R}^a a + R^d d^h + R^m m^h  +{s}_{\mathrm{max}}} \nonumber \\
    & \hspace{5cm} + \frac{1-p^\epsilon}{\bar{R}^a a + R^d d^h +R^m m^h + {s}_{\mathrm{min}}} \bigg ] \nonumber \\
    & \hspace{4cm} +\ubar{R}^a (1-p^a) \bigg [ \frac{p^\epsilon}{\ubar{R}^a a + R^d d^h + R^m m^h +{s}_{\mathrm{max}}} \nonumber \\
    & \hspace{5cm} + \frac{1-p^\epsilon}{\ubar{R}^a a + R^d d^h +R^m m^h + {s}_{\mathrm{min}}} \bigg ] \bigg \}, \label{foca_cbdc} \tag{C4}\\
    d^h: \qquad &\frac{1}{y-a-d^h-m^h} = \beta R^d \bigg \{ p^a \bigg [ \frac{p^\epsilon}{\bar{R}^a a + R^d d^h + R^m m^h +{s}_{\mathrm{max}}} \nonumber \\
    & \hspace{5cm} + \frac{1-p^\epsilon}{\bar{R}^a a + R^d d^h + R^m m^h + {s}_{\mathrm{min}}} \bigg ] \nonumber \\
    & \hspace{4cm} + (1-p^a) \bigg [ \frac{p^\epsilon}{\ubar{R}^a a +R^d d^h +{s}_{\mathrm{max}}} \nonumber \\
    & \hspace{5cm} + \frac{1-p^\epsilon}{\ubar{R}^a a +R^d d^h + R^m m^h + {s}_{\mathrm{min}}} \bigg ] \bigg \} \nonumber \\
    & \hspace{3.5cm} + \gamma \bigg [ \frac{(d^h)^{-\sigma}}{(d^h)^{1-\sigma} + \lambda(m^h)^{1-\sigma}} \bigg ], \label{focdh_cbdc} \tag{C5}\\
    m^h: \qquad &\frac{1}{y-a-d^h-m^h} = \beta R^m \bigg \{ p^a \bigg [ \frac{p^\epsilon}{\bar{R}^a a + R^d d^h + R^m m^h +{s}_{\mathrm{max}}} \nonumber \\
    & \hspace{5cm} + \frac{1-p^\epsilon}{\bar{R}^a a + R^d d^h + R^m m^h + {s}_{\mathrm{min}}} \bigg ] \nonumber \\
    & \hspace{4cm} + (1-p^a) \bigg [ \frac{p^\epsilon}{\ubar{R}^a a +R^d d^h +{s}_{\mathrm{max}}} \nonumber \\
    & \hspace{5cm} + \frac{1-p^\epsilon}{\ubar{R}^a a +R^d d^h + R^m m^h + {s}_{\mathrm{min}}} \bigg ] \bigg \} \nonumber \\
    & \hspace{3.5cm} + \gamma \lambda \bigg [ \frac{(m^h)^{-\sigma}}{(d^h)^{1-\sigma} + \lambda(m^h)^{1-\sigma}} \bigg ] \label{focmh} \tag{C6},
\end{align}
and the LFL agent's first-order conditions for deposits and CBDC are, respectively:
\begin{align}
    d^l: \qquad &\frac{1}{y-d^l-m^l} = \beta R^d \bigg [ \frac{p^\epsilon}{R^d d^l + R^m m^l +{s}_{\mathrm{max}}} + \frac{1-p^\epsilon}{R^d d^l + {s}_{\mathrm{min}}} \bigg ] \nonumber \\
    & \hspace{3.5cm} + \gamma \bigg [ \frac{(d^l)^{-\sigma}}{(d^l)^{1-\sigma} + \lambda(m^l)^{1-\sigma}} \bigg ] \label{focdl_cbdc} \tag{C7}, \\
    m^l: \qquad &\frac{1}{y-d^l-m^l} = \beta R^m \bigg [ \frac{p^\epsilon}{R^d d^l + R^m m^l +{s}_{\mathrm{max}}} + \frac{1-p^\epsilon}{R^d d^l + {s}_{\mathrm{min}}} \bigg ] \nonumber \\
    & \hspace{3.5cm} + \gamma \lambda \bigg [ \frac{(m^l)^{-\sigma}}{(d^l)^{1-\sigma} + \lambda(m^l)^{1-\sigma}} \bigg ] \label{focml} \tag{C8}.
\end{align}

\subsection{Risky asset calibration} \label{sec:app:cal}
I calibrate risky-asset returns and probabilities using the binomial tree security pricing model [see, e.g., \textcite{Petters2016}].

I use market data for a representative mix of risky assets, considering a high return $\bar{r}^a = 1.09$ and a low return $\ubar{r}^a = 0.6$. The high return reflects the average for best-performing years in recent periods [\textcite{FED2021}]. The low return reflects the worst-case scenario if the agent exits the market at the lowest point of the Global Financial Crisis.

The probabilities for these returns are calibrated to match an annual equity premium of approximately $6\%$ consistent with recent estimates [\textcite{Damodaran2024}]. The equity premium is defined as:
\begin{align}
    \mathrm{Equity} \, \mathrm{premium} = \mathbb{E}[r^a] - r^d, \label{ypa} \tag{C9}
\end{align}
where $r^d$ is the annual risk-free return equal to $1.01$. The stochastic returns on risky assets follow:
\begin{align} \label{yra} \tag{C10}
  r^a =
    \begin{cases}
      \bar{r}^a = 1.09 & \text{with probability $f^a = 0.95$},\\
      \ubar{r}^a = 0.6 & \text{with probability $1-f^a = 0.05$}.
    \end{cases}       
\end{align}

Over the two periods of equal duration $T$ years, the compounded risky asset returns follow a random multiplicative process:
\begin{align} \tag{C11}
    R^a = 
    \begin{cases}
     (\bar{r}^a)^{(T-n)} (\ubar{r}^a)^n & \text{with probability $P = \binom{T}{n} (f^a)^{(T-n)} (1-f^a)^n$},
    \end{cases}
\end{align}
where $n$ represents the number of crisis years. For example, assuming a period duration of $20$ years and considering probabilities as in equation (\ref{yra}), the case of one crisis year has a return on the risky assets equal to
\begin{align}
    R^a_1 = (1.09)^{19} (0.6)^1 = 3.08, \tag{C12}
\end{align}
and happens with probability
\begin{align}
    P_1 = \binom{20}{1} (0.95)^{19} (0.05)^1 = 0.38. \tag{C13}
\end{align}
The possible outcomes over the period are reported in Table \ref{cal_risky}.

\begin{table}[!htbp]
\centering
\caption{Stochastic risky asset return over the period}
\begin{threeparttable}
    \begin{tabular*}{0.6\linewidth}{@{\extracolsep{\fill}}l*{3}{c}}
    \toprule \toprule
    \multicolumn{1}{c}{$n$}&\multicolumn{1}{c}{$P_n$}&\multicolumn{1}{c}{$R^a_n$}\\
    \midrule 
    $0$ & $0.358$ & $5.60$ \\
    $1$ & $0.377$ & $3.08$ \\
    $2$ & $0.188$ & $1.69$ \\
    \midrule
    $3$ & $0.059$ & $0.93$ \\
    $4$ & $0.013$ & $0.51$ \\
    $5$ & $0.002$ & $0.28$ \\
    $\dots$ & $\dots$ & $\dots$ \\
    \bottomrule
    \end{tabular*}
\end{threeparttable}
\label{cal_risky}
\end{table}

To simplify, I bin the outcomes into two events: a high-return event and a low-return event. The high return occurs with the probability
\begin{align}
    p^a = P_0 + P_1 + P_2 = 0.92, \tag{C14}
\end{align}
and gives a weighted average return of
\begin{align}
    \bar{R}^a = \frac{P_0 R^a_0 + P_1 R^a_1 + P_2 R^a_2}{P_0 + P_1 + P_2} = 3.77. \tag{C15}
\end{align}
The low return event occurs with a probability approximately equal to $(1-p^a) = 0.08$ and gives a weighted average return of
\begin{align}
    \ubar{R}^a = \frac{P_3 R^a_3 + P_4 R^a_4 + P_5 R^a_5}{P_3 + P_4 + P_5} = 0.83. \tag{C16}
\end{align}
\newpage

\section{Additional figures} \label{sec:app:addfig}
\renewcommand{\thetable}{D.\arabic{table}}
\renewcommand{\thefigure}{D.\arabic{figure}}
\setcounter{table}{0}
\setcounter{figure}{0}

\begin{figure}[!htbp]
\centering
\caption{Portfolio allocation for the low-financially literate agent with stochastic income as a function of CBDC return}
\begin{minipage}{\textwidth}
    \centering
    \makebox[\textwidth]{\includegraphics[width=0.8\textwidth]{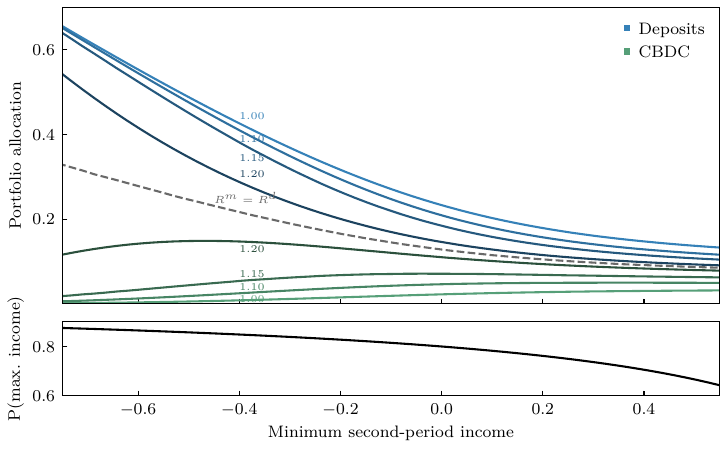}}
\end{minipage}
\vskip 0.5em
\begin{minipage}{\textwidth}
    \footnotesize
    \textit{Note}: This figure shows the portfolio allocation in the CBDC economy for the low-financially literate agent with stochastic income. Model solutions are presented for different values of CBDC return, $R^m$, equal to $1, 1.10, 1.15$, and $1.20$, with light color lines corresponding to lower returns. The dashed line indicates the case for $R^m = R^d \simeq 1.22$, where the allocations for CBDC and deposits coincide. Values are relative to the first-period endowment $y=1$.
\end{minipage}
\label{fig:portstocrm}
\end{figure}

\end{document}